\documentclass[journal]{IEEEtran}
\usepackage{color}
\usepackage{enumitem}

\usepackage{cite}

\ifCLASSINFOpdf
  \usepackage[pdftex]{graphicx}
  \graphicspath{{./fig/}}
\else
\fi
\usepackage{amsmath,amssymb}

\usepackage{stfloats}
\begin{document}
%
\title{CRLB Approaching Pilot-aided Phase and Channel Estimation Algorithm in MIMO Systems with Phase Noise and Quasi-Static Channel Fading}
%
%
%

\author{Yiming~Li,
        Zhouyi~Hu,~\IEEEmembership{Member,~IEEE,}
        and~Andrew~D.~Ellis

\thanks{Manuscript received May 00, 2021; revised May 00, 2021. Research supported by EPSRC under grant number EP/T009047/1, and the European Union’s Horizon 2020 research and innovation programme under the Marie Skłodowska-Curie grant agreement No. 713694.}
\thanks{Y. Li, Z. Hu, and A. Ellis are with Aston Institute of Photonic Technology, Aston University, Birmingham, B4 7ET, UK (e-mail: y.li70@aston.ac.uk; z.hu6@aston.ac.uk; andrew.ellis@aston.ac.uk).}
}

%
%

\markboth{IEEE Transactions on signal processing,~Vol.~00, No.~0, August~2021}%
{Shell \MakeLowercase{\textit{et al.}}: Bare Demo of IEEEtran.cls for IEEE Journals}
%



\maketitle

\begin{abstract}
This paper derives a novel pilot-aided phase and channel estimation algorithm for multiple-input multiple-output (MIMO) systems with phase noise and quasi-static channel fading. Our novel approach allows, for the first time, carrier phase estimation and recovery to be performed before full channel estimation. This in turn enables the channel estimation to be calculated using the whole frame, significantly improving its accuracy. The proposed algorithm is a sequential combination of several linear algorithms, which greatly reduces the computational complexity. Moreover, we also derive, for the first time, the Cramér–Rao lower bound (CRLB) for a MIMO system, where phase noise is estimated using only angular information. Our numerical results show that the performance of our phase estimation algorithm is close to the proposed CRLB. Moreover, when compared with the conventional Kalman based algorithms, our proposed algorithm significantly improves the system BER performance.

\end{abstract}


%
\IEEEpeerreviewmaketitle

\section{Introduction}
\label{sec:in}
%
%
%
%

\subsection{Motivation}
\IEEEPARstart{M}{ultiple}-input multiple-output (MIMO) systems can significantly enhance the performance of both wireless and optical communication systems in terms of capacity, outage performance, and bit error rate (BER) performance \cite{telatar1999capacity,tse2005fundamentals,stuart2000dispersive,sleiffer201273,wilson2003optical}. In general, the theoretical performance limit of MIMO systems are based on the assumption of perfect channel state information at the receiver (CSIR). However, practical MIMO systems suffer from imperfect CSIR, which may significantly deteriorate the performance from the theoretical limit. Therefore, accurate and efficient estimation algorithms are important for practical systems.

Phase noise, which is induced by oscillator imperfections and varies from symbol to symbol, is one of the most important detrimental effects in coherent communication systems. In MIMO systems, the influence of phase noise can be more pronounced when independent oscillators are connected to different transmit and receive antennas. This scenario is well motivated when each transmit and receive antennas are far away from each other \cite{bohagen2007design}. Recently, a lot of relevant phase and channel estimation applications has been investigated for different scenarios such as joint estimation and detection problem in orthogonal frequency-division multiplexing (OFDM) relay systems \cite{wang2015channel}, phase estimation problem in space division multiplexing multicore fibre transmissions \cite{alfredsson2018pilot}, and joint channel and location estimation for massive MIMO systems \cite{zheng2020joint}.

Although extensive research on phase noise estimation has been done on single-input single-output (SISO) systems \cite{noels2005carrier,zhao2006novel,Ip2007p2675}, they can not be directly applied to MIMO systems, where the received signal may be deteriorated by multiple phase noise parameters.
Therefore, the joint phase and channel estimation for MIMO systems with phase noise is of particular interest and has been investigated by several researches \cite{mehrpouyan2012joint,nasir2013phase,reggiani2018extended,alfredsson2019iterative,krishnan2015algorithms}. In \cite{mehrpouyan2012joint}, extended-Kalman filter (EKF) was proposed for phase estimation. And an improved version of extended-Kalman smoother (EKS) is proposed in \cite{nasir2013phase} to better estimate the phase estimation performance by EKF-EKS algorithm. Recently, foward-backward EKF and iterative phase compensation using EKS are also considered to further improve the system performance \cite{reggiani2018extended,alfredsson2019iterative}. However, the EKF/EKS based algorithms have to estimate the phase information \textit{after} channel estimation.
Unfortunately, the performance of the conventional least-squares (LS) channel estimation algorithm is deteriorated by phase noise and additive white Gaussian noise (AWGN), which leads to inaccurate channel estimation \cite{mehrpouyan2012joint}. As a result, at the hard-decision forward error corrction (HD-FEC) limit of \(4.7 \times 10^{-3}\) (\(6.25\%\) overhead) \cite{zhang2014staircase,alvarado2015replacing}, there is an approximately 3~dB signal-to-noise ratio (SNR) penalty in the BER performance when compared to the theoretical limit with perfect phase and channel estimation \cite{mehrpouyan2012joint,nasir2013phase}.

Another approach for better estimation accuracy and system performance is to \textit{jointly} estimate the phase and channel by applying maximum a posteriori (MAP) algorithms, which outperforms the EKF/EKS approach and quantifies slight degradation when compared to Cramér–Rao Lower Bound (CRLB)  \cite{nasir2013phase}. Moreover, different simplified algorithms from MAP scheme has been discussed in \cite{krishnan2015algorithms}. However, the computational complexity of the MAP-based algorithms grows exponentially, which is not suitable for practical MIMO systems with a relatively large number of transmit and receive antennas.

\subsection{Contributions}
In this paper, a new pilot-aided phase and channel estimation algorithm is proposed for MIMO systems with phase noise. The proposed algorithm significantly improves the phase and channel estimation accuracy and reduces the SNR penalty when compared to the Kalman based approaches. Moreover, the algorithm significantly reduces the computational complexity of the estimation process when compared with all existing algorithms. The main contributions of this paper are summarized as follows.
\begin{enumerate}
  
  \item{A new algorithm is proposed to estimate the phase and channel information. Unlike the existing algorithms, the proposed algorithm splits the phase and channel estimation into its phase invariant amplitude part, estimates and recovers the phase noise, and then estimates the full (complex) channel matrix. 
  By doing so, the interaction between phase noise and channel estimation reduces to a quasi-static phase shift, which cancels out when the overall phase and channel estimation is computed.
  Moreover, the proposed algorithm allows the phase estimation to be performed \textit{before} full channel estimation, and} enables the averaging process for channel estimation over the whole frame. As a result, this algorithm can significantly improve the phase and channel estimation accuracy, and reduce the SNR penalty of MIMO decoders. At the HD-FEC limit of \(4.7 \times 10^{-3}\) \cite{alvarado2015replacing}, the proposed algorithm has an SNR penalty of approximately 0.5~dB compared to the ideal case with perfect phase and channel estimation, which is more than 2~dB lower than the known EKF/EKS algorithms.
  
  \item{Under the assumption of estimating the phase information from the angular terms of the observed data, a new CRLB is derived. Although this CRLB is higher when compared to the ultimate CRLB in the high SNR region \cite{nasir2013phase}, it is preferable for the practical situation of estimating the phase information from angular terms of the observed data. The numerical results show that the performance of the proposed algorithm is very close to the CRLB.}
  
  \item{The algorithm is a sequential combination of several linear algorithms, which greatly reduces the computational complexity of the phase and channel estimation process and makes it feasible for practical commercial systems. To the best of our knowledge, the proposed algorithm has the lowest computational complexity among all the existing algorithms.}
\end{enumerate}

\subsection{Organization}
The remainder of this paper is organized as follows: the system model is given in Section~\ref{sec:sm}, Section~\ref{sec:ea} proposes the novel phase and channel estimation algorithm, Section~\ref{sec:crlb} derives the CRLB from the angular terms of the observed data, Section~\ref{sec:nr} provides the numerical results of the proposed algorithm in the MIMO systems, and Section~\ref{sec:c} summarizes the key advantages of the proposed algorithm.

\subsection{Notations}

Unless otherwise specified, boldface capital letters, e.g. \(\bf{X}\), are used for matrices. Boldface small letters, e.g. \(\bf{x}\), are used for vectors. More specifically, \({\bf{x}}_{m,:}\) is used for the \(m^{th}\) row vector of \(\bf{X}\), and \({\bf{x}}_{:,m}\) is used for the \(m^{th}\) column vector of \(\bf{X}\). Moreover, regular small letters with subscript indices, e.g. \(x_m\) and \(x_{m,n}\), represents the \(m^{th}\) element of \(\bf{x}\) and the element at the \(m^{th}\) row and \(n^{th}\) column of \(\bf{X}\), respectively.
\({{\bf{I}}_{X}}\) denotes the \({{X} \times {X}}\) identity matrix.
\({{\bf{0}}_{X \times Y}}\) and \({{\bf{1}}_{{X} \times {Y}}}\) denotes the \({{X} \times {Y}}\) all zero and all one matrices, respectively.
\({\left(  \cdot  \right)^H}\), \({\left(  \cdot  \right)^T}\), \({\left(  \cdot  \right)^ + }\), \({\left(  \cdot  \right)^ * }\), and \({\left|  \cdot  \right|}\) denote the Hermitian transpose, transpose, Moore–Penrose inverse, element-wise conjugate, and element-wise absolute value operators, respectively.
\(\left\|  \cdot  \right\|\) denotes the Euclidean norm of a vector.
\( \odot \) and \( \oslash \) denote the element-wise product and element-wise division of two matrices, respectively.
\(\max \left( {{\bf{X}},{\bf{Y}}} \right)\) denotes the element-wise maximum operator of two matrices \({\bf{X}}\) and \({\bf{Y}}\).
\(\left\lceil  x  \right\rceil \) denotes the ceiling function which returns the smallest integer greater than or equal to \(x\).
\(\angle \left(  \cdot  \right)\) denotes the element-wise phase angle of complex matrices.
\({\rm{diag}}\left( {\bf{x}} \right)\) denotes a diagonal matrix, and its diagonal elements are given by \(\bf{x}\).
And \({\rm{diag}}\left( {\bf{X}} \right)\) denotes a vector which contains the diagonal elements of \(\bf{X}\).
\(\Re \left\{  \cdot  \right\}\) and \(\Im \left\{  \cdot  \right\}\) denote the real part and the imaginary part of a complex number, respectively.
\({\text{E}}\left(  \cdot  \right)\) denotes the expected value of a variable.
Finally, \({\cal N}\left( {\mu,\sigma^2} \right)\) and \({\cal {CN}}\left( {\mu,\sigma^2} \right)\) denote real and complex Gaussian distributions with mean \(\mu\) and variance \(\sigma^2\), respectively.

\section{System Model}
\label{sec:sm}

\begin{figure}[htb]
  \centering
  \includegraphics{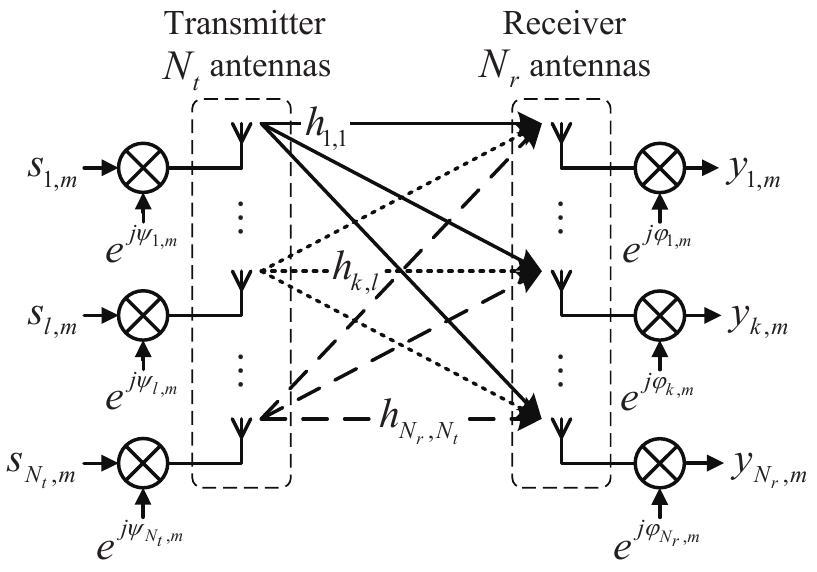}
  \caption{Schematic diagram for MIMO systems}
  \label{fig:schematic}
  \end{figure}

As shown in Fig.~\ref{fig:schematic}, a point-to-point MIMO system with \(N_t\) transmit antennas and \(N_r\) receive antennas is considered. In Fig.~\ref{fig:schematic}, \(s_{l,m}\) represents the transmitted symbol at the \(l^{th}\) transmitter and the \(m^{th}\) time interval. \(y_{k,m}\) represents the received signal at the \(k^{th}\) receiver and the \(m^{th}\) time interval. \(h_{k,l}\) is the channel gain between the \(k^{th}\) receiver and the \(l^{th}\) transmitter. \(\psi_{l,m}\) represents the phase noise at the \(l^{th}\) transmitter and the \(m^{th}\) time interval. And \(\varphi_{l,m}\) represents the phase noise at the \(k^{th}\) receiver and the \(m^{th}\) time interval.

\begin{figure}[htb]
  \centering
  \includegraphics{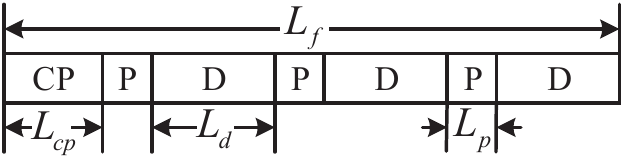}
  \caption{Frame structure of the transmitted data in MIMO systems. CP: Cyclic Prefix, P: Pilot, D: Data.}
  \label{fig:frame}
  \end{figure}

As shown in Fig.~\ref{fig:frame}, data symbols are transmitted as frames. The frame length is \(L_f\). In each frame, there is a cyclic prefix (CP) of length \(L_{cp}\). After the CP, \(L_p\) consecutive \textit{pilot} symbols (a pilot group) are inserted every \(L_d\) \textit{data} symbols. The cell length (\(L_c\)), cell number (\(N_c\)), and pilot rate (\(R_p\)) of a frame are defined as
  \begin{equation}\label{equ:1}
  \left\{ {\begin{aligned}
    L_c &= {L_p} + {L_d}\\
    N_c &= \frac{{{L_f} - {L_{cp}}}}{{{L_p} + {L_d}}}\\
    R_p &= \frac{{{L_p}}}{{{L_p} + {L_d}}}
  \end{aligned}} \right..
  \end{equation}

In this paper, the following set of assumptions are adopted:

\renewcommand{\labelenumi}{A\arabic{enumi}.}
\begin{enumerate}[itemindent=1em]
  \item {The pilots are assumed to be \textit{known} at the receiver. Moreover, it is assumed that all the transmitters simultaneously transmit mutually orthogonal and element-wise normalized pilots of length \(L_p = N_t\) in one pilot group to the receiver (e.g. the element-wise normalized discrete Fourier transform (DFT) matrix).}
  
  \item {As shown in Fig.~\ref{fig:schematic}, the oscillators of different transmit and receive antennas are assumed to be independent, which is a generalized case of different kinds of practical systems \cite{bohagen2007design,hadaschik2005improving}.}
  \item {The channels considered in this paper are assumed to be quasi-static and frequency-flat fading channels. Therefore, the channel matrix \({\bf{H}}\) remains constant in a frame and there is no inter-symbol interference (ISI) in the system.}
  \item {The phase noise is modeled as Wiener process \cite{demir2000phase,chorti2006spectral,Ip2007p2675}. Considering the fact that the phase noise is much slower than the symbol rate, it is assumed that the phase is static in one symbol period but varying from symbol to symbol \cite{mehrpouyan2012joint, wang2015channel,pitarokoilis2014uplink,zhang2020downlink}. Moreover, the intra-pilot-group phase noise refers to the phase noise within one pilot group, and the inter-pilot-group phase noise refers to the phase noise between different pilot groups throughout this paper.}
  \item {Perfect SNR estimation, timing recovery, frame and bit synchronization, and frequency estimation are assumed, which can be achieved by standard estimation and synchronization algorithms \cite{zhao2006novel, schmidl1997robust, simon2001phase, hadaschik2005improving, Proakis2008communications}.}
\end{enumerate}

In the above mentioned system, the discrete-time baseband received signal of the system can be described as
  \begin{equation}\label{equ:2}
  {{\bf{y}}_{:,m}} = {{\bf{\Phi }}_m}{\bf{H}}{{\bf{\Psi }}_m}{{\bf{s}}_{:,m}} + {{\bf{n}}_{:,m}},
  \end{equation}
where 

\(m\) represents the \(m^{th}\) time interval,

\({{\bf{y}}_{:,m}} = {\left[ {{y_{1,m}}, \cdots ,{y_{{N_r},m}}} \right]^T}\) is the received signal vector,

\({{\mathbf{\Phi }}_m} = {\text{diag}}\left( {{e^{j{\varphi _{1,m}}}}, \cdots ,{e^{j{\varphi _{{N_r},m}}}}} \right)\) is the diagonal phase matrix of the receiver oscillators,

\({\bf{H}} = \left[ {\begin{array}{*{20}{c}}
{{h_{1,1}}}& \cdots &{{h_{1,{N_t}}}}\\
 \vdots & \ddots & \vdots \\
{{h_{{N_r},1}}}& \cdots &{{h_{{N_r},{N_t}}}}
\end{array}} \right]\) is the channel matrix, 

\({{\mathbf{\Psi }}_m} = {\text{diag}}\left( {{e^{j{\psi _{1,m}}}}, \cdots ,{e^{j{\psi _{{N_t},m}}}}} \right)\) is the diagonal phase matrix of the transmitter oscillators,

\({{\bf{s}}_{:,m}} = {\left[ {{s_{1,m}}, \cdots ,{s_{{N_t},m}}} \right]^T}\) is the transmitted signal vector with normalized power,

\({{\bf{n}}_{:,m}} = {\left[ {{n_{1,m}}, \cdots ,{n_{{N_r},m}}} \right]^T}\) is the independent and identically distributed (i.i.d.) circularly-symmetric complex AWGN vector, and \({n_{k,m}} \sim {\cal C}{\cal N}\left( {0,\sigma _n^2} \right)\).

Following assumption (A4), the phase of the oscillators for different transmit and receive antennas can be modelled as Wiener process. And \({\varphi _{k,m}}\) and \({\psi _{l,m}}\) are then given by
  \begin{equation}\label{equ:3}
  {\varphi _{k,m}} = {\varphi _{k,m - 1}} + \Delta {\varphi _{k,m}},
  \end{equation}
  \begin{equation}\label{equ:4}
  {\psi _{l,m}} = {\psi _{l,m - 1}} + \Delta {\psi _{l,m}},
  \end{equation}
where \(\Delta {\varphi _{k,m}} \sim {\cal N}\left( {0,\sigma _{\Delta \varphi }^2} \right)\) and \(\Delta {\psi _{l,m}} \sim {\cal N}\left( {0,\sigma _{\Delta \psi }^2} \right)\) represents the phase innovations of the Wiener process at the \(k^{th}\) receive and the \(l^{th}\) transmit oscillator, respectively, at the \(m^{th}\) symbol period. \(\sigma _{\Delta \varphi }^2\) and \(\sigma _{\Delta \psi }^2\) represents the variance of the phase noise innovation at the transmitters and the receivers, respectively.


\textit{Remark 1:} Note the fact that \({\bf{H}}\) is a matrix with rank \(N_t\) in a general MIMO system with \(N_r \ge N_t\), a feasible phase and channel estimation can not be obtained within one pilot group if \(L_p < N_t\) (there are insufficient equations to solve the channel matrix). In this case, although the phase and channel can be estimated by several pilot groups, the inter-pilot-group phase noise significantly contaminates the channel matrix estimation performance. On the other hand, 
the impact of intra-pilot-group phase noise on channel estimation increases as \(L_p\) increases (more phase wander during \(L_p\)), and so minimizing \(L_p\) is recommended.
Therefore, the optimal condition of \(L_p = N_t\) is important in assumption (A1).

\section{Estimation Algorithm}
\label{sec:ea}
\subsection{Basic Principles of the Estimation Algorithm}

Throughout our algorithm, \(pilots\) are used for estimating the phase and channel, while CP is only used for frame synchronization and frequency estimation in assumption (A5). Without loss of generality, the time index of the first symbol in the first pilot group is labelled as \(m=1\)  (the time indices of the CP are \(- L_{cp}+1 \le m \le 0\), which are neglected in the algorithm).

Following assumption (A3), a suitable estimation of channel matrix \(\bf{H}\) should exploit all the pilot information throughout a frame to reduce estimation error. However, if the channel estimation is performed before phase estimation, there is a trade-off between the AWGN and the phase noise, and the channel estimation can be severely contaminated by phase noise when using all the pilots. Therefore, a natural idea is to estimate and recover phase before channel estimation to remove the trade-off.

In order to estimate the phase of different transmit and receive antennas within a pilot group, a weighted linear least-squares (WLLS) estimator is proposed in this chapter. Moreover, an element-wise Wiener filter array is applied to further improve the phase estimation accuracy by exploiting the inter-pilot-group phase information.

It will be shown later in this chapter that the WLLS estimator needs the information of \(\left| {\bf{H}} \right|\) as weight coefficients. By applying conventional LS algorithm, the estimation of \(\left| {\bf{H}} \right|\) is possible in each pilot group. Because of the absolute operator, the phase contamination phenomenon can not influence the amplitude of the channel matrix. Therefore, it is possible to calculate the element-wise average of \(\left| {\bf{H}} \right|\) throughout all the pilots within a frame before phase recovery.

\begin{figure}[htb]
  \centering
  \includegraphics[width=1.8in]{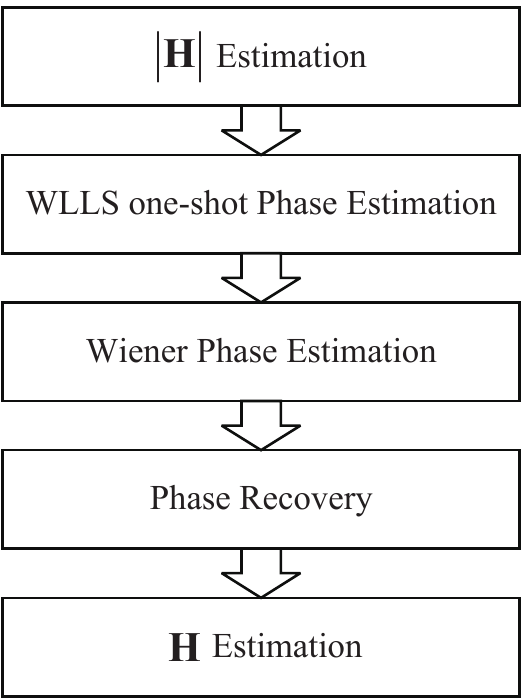}
  \caption{Structure of estimation algorithm.}
  \label{fig:AlgorithmStructure}
  \end{figure}

As a conclusion of the above mentioned reasons, the structure of our newly proposed algorithm is shown in Fig.~\ref{fig:AlgorithmStructure}.

It is worth noting that a quasi-static phase bias will occur in the WLLS estimation. This bias also exists (but in the negative direction) in the full (complex) channel estimation. Therefore, the bias can be cancelled out, which guarantees a feasible estimation of \({{\bf{\Phi }}_{{m}}}{\bf{H}}{{\bf{\Psi }}_{{m}}}\).

\textit{Remark 2:} The influence of the intra-pilot-group phase noise is neglectable in the low SNR region and very subtle in the high SNR region. Therefore, it is neglected throughout the proposed algorithm. The details of intra-pilot-group phase noise will be discussed in Sec.~\ref{sec:crlb}. And the numerical result of the performance degradation will be discussed in Sec.~\ref{sec:nr}. 
  
\subsection{Step 1: Estimation of Channel Amplitude}
It is straightforward to prove that the phase term \({{\bf{\Phi }}_m}\) and \({{\bf{\Psi }}_m}\) does not change the amplitude of the equivalent channel matrix \({{\bf{\Phi }}_m}{\bf{H}}{{\bf{\Psi }}_m}\). Therefore, the estimation of channel amplitude at the \(i^{th}\) pilot group can be given by conventional LS algorithm as \cite{mehrpouyan2012joint}
  \begin{equation}\label{equ:5}
  \left| {{{{\bf{\hat H}}}_i}} \right| = \left| {{{\bf{Y}}_i}{\bf{S}}_i^ + } \right|,
  \end{equation}
where \({{\bf{Y}}_i} = \left[ {{{\bf{y}}_{:,\left( {i - 1} \right) \times {L_c} + 1}}, \cdots ,{{\bf{y}}_{:,\left( {i - 1} \right) \times {L_c} + {N_t}}}} \right]\) and \({{\bf{S}}_i} = \left[ {{{\bf{s}}_{:,\left( {i - 1} \right) \times {L_c} + 1}}, \cdots ,{{\bf{s}}_{:,\left( {i - 1} \right) \times {L_c} + {N_t}}}} \right]\).

Following assumption (A1), \({{\bf{S}}_i}{{\bf{S}}_i^H} = {N_t}{{\bf{I}}_{N_t}}\). Therefore, the estimation of channel amplitude at the \(i^{th}\) pilot group can be
  \begin{equation}\label{equ:6}
  \left| {{{{\bf{\hat H}}}_i}} \right| = \frac{1}{{{N_t}}}\left| {{{\bf{Y}}_i}{\bf{S}}_i^H} \right|.
  \end{equation}
  
When the intra-pilot-group phase noise is neglected, and the reference phase is set at the middle of the \(i^{th}\) pilot group, the time index of the reference phase can be defined as
  \begin{equation}\label{equ:7}
    {m_i} \triangleq \left( {i - 1} \right) \times {L_c} + \left\lceil {{N_t}/2} \right\rceil.
  \end{equation}
And \eqref{equ:6} can be written as
  \begin{equation}\label{equ:8}
  \begin{aligned}
  \left| {{{{\bf{\hat H}}}_i}} \right| & \approx \frac{1}{{{N_t}}}\left| {\left( {{{\bf{\Phi }}_{{m_i}}}{\bf{H}}{{\bf{\Psi }}_{{m_i}}}{{\bf{S}}_i} + {{\bf{N}}_i}} \right){\bf{S}}_i^H} \right|\\
  & = \left| {{{\bf{\Phi }}_{{m_i}}}{\bf{H}}{{\bf{\Psi }}_{{m_i}}} + \frac{1}{{{N_t}}}{{\bf{N}}_i}{\bf{S}}_i^H} \right|\\
  & = \left| {{{\bf{\Phi }}_{{m_i}}}{\bf{H}}{{\bf{\Psi }}_{{m_i}}} + {{{{\bf{N}}_i'}}}} \right|,
  \end{aligned}
  \end{equation}
where \({{\bf{N}}_i} = \left[ {{{\bf{n}}_{:,\left( {i - 1} \right) \times {L_c} + 1}}, \cdots ,{{\bf{n}}_{:,\left( {i - 1} \right) \times {L_c} + {N_t}}}} \right]\) is the corresponding AWGN matrix. Moreover, \({{{\bf{S}}_i^H} \mathord{\left/
 {\vphantom {{{\bf{S}}_i^H} {\sqrt {{N_t}} }}} \right.
 \kern-\nulldelimiterspace} {\sqrt {{N_t}} }}\) is a unitary matrix. Therefore, \({\bf{N}}_i'\) is also an i.i.d. circularly-symmetric Gaussian matrix. And the element at the \(k^{th}\) row and \(l^{th}\) column of \({\bf{N}}_i'\) obeys the distribution \(n_{i,k,l}' \sim {\cal C}{\cal N}\left( {0,{{\sigma _n^2} \mathord{\left/
 {\vphantom {{\sigma _n^2} {{N_t}}}} \right.
 \kern-\nulldelimiterspace} {{N_t}}}} \right)\).

Noting the i.i.d. property of \({\bf{N}}_i'\),
the element-wise squaring of channel amplitude estimation can be calculated over the whole frame as
  \begin{equation}\label{equ:9}
  {\left| {{\mathbf{\hat H}}} \right|_{sq}} = \frac{1}{{{N_c}}}\sum\limits_{i = 1}^{{N_c}} {\left( {\left| {{{{\mathbf{\hat H}}}_i}} \right| \odot \left| {{{{\mathbf{\hat H}}}_i}} \right|} \right)}  - \sigma _n^2{{\mathbf{1}}_{{N_r} \times {N_t}}}.
  \end{equation}

Finally, the estimation of channel amplitude over the whole frame can be calculated as
  \begin{equation}\label{equ:10}
  \left| {{\mathbf{\hat H}}} \right| = \sqrt {\max \left( {{{\mathbf{0}}_{{N_r} \times {N_t}}},{{\left| {{\mathbf{\hat H}}} \right|}_{sq}}} \right)}.
  \end{equation}
The max operator in \eqref{equ:10} is to prevent the estimated value in \eqref{equ:9} from accidentally being less than 0 when \({h_{i,j}} \to 0\).

\textit{Remark 3:} A conventional estimate of channel amplitude would be \(\left| {{\bf{\hat H}}} \right| = \left| {\frac{1}{{{N_c}}}\sum\limits_{i = 1}^{{N_c}} {{{{\bf{\hat H}}}_i}} } \right|\) or \(\left| {{\bf{\hat H}}} \right| = \frac{1}{{{N_c}}}\sum\limits_{i = 1}^{{N_c}} {\left| {{{{\bf{\hat H}}}_i}} \right|} \). However, in the first case, the estimation is corrupted by inter-pilot-group phase noise, reducing the accuracy for high phase noise, whilst for the latter the absolute value of the observable is dominated by AWGN at low SNR (consider the case of \({h_{k,l}} = 0\), this estimation gives a result which depends on the AWGN and is larger than 0). Both of these issues are resolved by adopting (9) and (10).

\textit{Remark 4:} Online version of \eqref{equ:9} is preferable in real-time systems to reduce latency. And the element-wise squaring of \(\left| {{\bf{\hat H}}} \right|\) at the \(i^{th}\) pilot group can be calculated as
  \begin{equation}\label{equ:11}
  \begin{aligned}
  {\left| {{\mathbf{\hat H}}} \right|_{sq\left( i \right)}} = & \left( {1 - K} \right){\left| {{\mathbf{\hat H}}} \right|_{sq\left( {i - 1} \right)}} \hfill \\
  + & K\left( {\left| {{{{\mathbf{\hat H}}}_i}} \right| \odot \left| {{{{\mathbf{\hat H}}}_i}} \right| - \sigma _n^2{{\mathbf{1}}_{{N_r} \times {N_t}}}} \right) \hfill ,\\ 
  \end{aligned}
  \end{equation}
where \(K\) is the updating factor which should be less than the coherent time of the channel.

\subsection{Step 2: Weighted Linear Least-Squares Phase Estimation}
Similar to \eqref{equ:8}, the following approximation holds
  \begin{equation}\label{equ:13}
    {{\mathbf{\hat H}}_i} = \frac{1}{{{N_t}}}{{\mathbf{Y}}_i}{\mathbf{S}}_i^H \approx {{\mathbf{\Phi }}_{{m_i}}}{\mathbf{H}}{{\mathbf{\Psi }}_{{m_i}}} + {{\mathbf{N_i'}}}.
  \end{equation}

Define the angular term of \({\mathbf{\hat H}}_i\) and \(\mathbf{H}\) as
  \begin{equation}\label{equ:14}
  {{\mathbf{A}}_i} \triangleq \angle {{\mathbf{\hat H}}_i},
  \end{equation}
  \begin{equation}\label{equ:15}
  {\bf{\Theta }} \triangleq \angle {\bf{H}}.
  \end{equation}
The element at the \(k^{th}\) row and \(l^{th}\) column of \({\bf{A}}_i\) can be represented as
  \begin{equation}\label{equ:16}
  \begin{aligned}
    {a_{i,k,l}} & \approx \angle \left( {{e^{j{\varphi _{k,{m_i}}}}}\left| {{h_{k,l}}} \right|{e^{j{\theta _{k,l}}}}{e^{j{\psi _{l,{m_i}}}}} + {{n}_{i,k,l}'}} \right)\\
    & = \angle \left( {\left| {{h_{k,l}}} \right|{e^{j\left( {{\varphi _{k,{m_i}}} + {\psi _{l,{m_i}}} + {\theta _{k,l}}} \right)}} + \left| {{h_{k,l}}} \right|\frac{{{{n}_{i,k,l}'}}}{{\left| {{h_{k,l}}} \right|}}} \right)\\
    & \approx {\varphi _{k,{m_i}}} + {\psi _{l,{m_i}}} + {\theta _{k,l}} + \frac{{{{n}''_{i,k,l}}}}{{\left| {{h_{k,l}}} \right|}},
  \end{aligned}
  \end{equation}
where \(\theta _{k,l}\) is the element at the \(k^{th}\) column and \(l^{th}\) row of \(\bf{\Theta}\), and \({n}_{i,k,l}'' \sim {\cal N}\left( {0,{{\sigma _n^2} \mathord{\left/ {\vphantom {{\sigma _n^2} {2{N_t}}}} \right. \kern-\nulldelimiterspace} {2{N_t}}}} \right)\) is the azimuthal component of \({n}_{i,k,l}'\). And the last approximation is based on the small noise assumption, which is also used in \cite{Ip2007p2675}. On the other hand, the angular term \(a_{:,k,l}\) given by \eqref{equ:16} is within the interval of \({\left[ { - \pi ,\pi } \right)}\), which is referred to as phase wrapping. Therefore, a standard phase unwrapping algorithm given by \cite{meyr1998digital} is necessary to unwrap each \({\bf{a}}_{:,k,l}\).

In order to eliminate phase ambiguity, it is intuitively satisfying to set one of the transmit oscillators as reference. This idea of setting reference phase is motivated by the results in \cite{stoica2001parameter, nasir2013phase, hadaschik2005improving}. Without loss of generality, the phase of the last transmit oscillator, \({\psi }_{N_t,m_i}\), is set as reference. and \eqref{equ:16} can be rewritten as
  \begin{equation}\label{equ:18}
  \begin{aligned}
    & {a_{i,k,l}} \\
    & \approx \left( {{\varphi _{k,{m_i}}} + {\psi _{{N_t},{m_i}}}} \right) + \left( {{\psi _{l,{m_i}}} - {\psi _{{N_t},{m_i}}}} \right) + {\theta _{k,l}} + \frac{{{{n}''_{i,k,l}}}}{{\left| {{h_{k,l}}} \right|}}\\
    & = {\beta _{k,{m_i}}} + {\beta _{N_r + l,{m_i}}} + {\theta _{k,l}} + \frac{{{{n}''_{i,k,l}}}}{{\left| {{h_{k,l}}} \right|}},
  \end{aligned}
  \end{equation}
where
  \begin{equation}\label{equ:19}
    {\beta _{q,m}} = \left\{ {\begin{aligned}
      {{\varphi _{q,m}} + {\psi _{{N_t},m}}} &{,\left( {1 \leqslant q \leqslant {N_r}} \right)} \\ 
      {{\psi _{q - {N_r},m}} - {\psi _{{N_t},m}}} &{,\left( {{N_r} + 1 \leqslant q \leqslant {N_r} + {N_t} - 1} \right)} ,
    \end{aligned}} \right. 
  \end{equation}
is one of the \(N_r + N_t - 1\) phase noise values correlated to the reference.

Define
  \begin{equation}\label{equ:20a}
    {\mathbf{C}} \triangleq \left[ {\begin{array}{*{20}{c}}
  {{{\mathbf{I}}_{{N_r}}}}&{{{{\mathbf{B}}}_1}} \\ 
   \vdots & \vdots  \\ 
  {{{\mathbf{I}}_{{N_r}}}}&{{{{\mathbf{B}}}_{{N_t} - 1}}} \\ 
  {{{\mathbf{I}}_{{N_r}}}}&{{{\mathbf{0}}_{{N_r} \times \left( {{N_t} - 1} \right)}}} 
  \end{array}} \right],
  \end{equation}
where \({{\bf{B}}_i} = \left[ {{{\bf{0}}_{{N_r} \times \left( {i - 1} \right)}},{{\bf{1}}_{{N_r} \times 1}},{{\bf{0}}_{{N_r} \times \left( {{N_t} - 1 - i} \right)}}} \right]\).
In order to enable matrix calculation, \({\bf{A}}_i\) is rearranged as
  \begin{equation}\label{equ:20b}
  {{\boldsymbol{\alpha }}_i} = {\left[ {{\mathbf{a}}_{i,:,1}^T, \cdots ,{\mathbf{a}}_{i,:,{N_t}}^T} \right]^T}.
  \end{equation}
Then \eqref{equ:18} can be rewritten in vector form as
  \begin{equation}\label{equ:20}
  {{\boldsymbol{\alpha }}_i} = {\mathbf{C}}{{\boldsymbol{\beta }}_{:,{m_i}}} + \left[ {\begin{array}{*{20}{c}}
  {{{\boldsymbol{\theta }}_{:,1}}} \\ 
   \vdots  \\ 
  {{{\boldsymbol{\theta }}_{:,{N_t}}}} 
\end{array}} \right] + \left[ {\begin{array}{*{20}{c}}
  {{{{\mathbf{n}}}_{i,:,1}''}} \\ 
   \vdots  \\ 
  {{{{\mathbf{n}}}_{i,:,{N_t}}''}} 
\end{array}} \right] \oslash \left[ {\begin{array}{*{20}{c}}
  {\left| {{{\mathbf{h}}_{:,1}}} \right|} \\ 
   \vdots  \\ 
  {\left| {{{\mathbf{h}}_{:,{N_t}}}} \right|} 
\end{array}} \right].
  \end{equation}
It is easy to verify that \(\bf{C}\) is a matrix with rank \(N_r+N_t-1\) (full rank) \cite{nasir2013phase}, which enables the estimation of \({{\boldsymbol{\beta }}_{:,{m_i}}}\).

Considering the fact that the LS algorithm requires i.i.d. noise on different observed data, if we assume \eqref{equ:10} has perfect estimation on the channel amplitude, \eqref{equ:20} can be modified by \eqref{equ:10} as
  \begin{equation}\label{equ:21}
    {{\boldsymbol{\alpha }}_i'} = {\bf{C}}'{{\boldsymbol{\beta }}_{:,{m_i}}} + \left[ {\begin{array}{*{20}{c}}
    {\left| {{{{\bf{\hat h}}}_{:,1}}} \right|}\\
     \vdots \\
    {\left| {{{{\bf{\hat h}}}_{:,{N_t}}}} \right|}
    \end{array}} \right] \odot \left[ {\begin{array}{*{20}{c}}
    {{{\boldsymbol{\theta }}_{:,1}}}\\
     \vdots \\
    {{{\boldsymbol{\theta }}_{:,{N_t}}}}
    \end{array}} \right] + \left[ {\begin{array}{*{20}{c}}
    {{{{\bf{n}}}_{i,:,1}''}}\\
     \vdots \\
    {{{{\bf{n}}}_{i,:,{N_t}}''}}
    \end{array}} \right],
  \end{equation}
where
  \begin{equation}\label{equ:22}
    {{{\boldsymbol{\alpha }}_i'} = {{\left[ {\left| {{\bf{\hat h}}_{:,1}^T} \right|, \cdots ,\left| {{\bf{\hat h}}_{:,{N_t}}^T} \right|} \right]}^T} \odot {{\boldsymbol{\alpha }}_i}},
  \end{equation}
  \begin{equation}\label{equ:23}
    {{\bf{C}}' = \underbrace {\left[ {\begin{array}{*{20}{c}}
    {\left| {{{{\bf{\hat h}}}_{:,1}}} \right|}& \cdots &{\left| {{{{\bf{\hat h}}}_{:,1}}} \right|}\\
     \vdots & \cdots & \vdots \\
    {\left| {{{{\bf{\hat h}}}_{:,{N_t}}}} \right|}& \cdots &{\left| {{{{\bf{\hat h}}}_{:,{N_t}}}} \right|}
    \end{array}} \right]}_{{\rm{repeat}}\ {N_r} + {N_t} - 1\ {\rm{times}}} \odot {\bf{C}}}.
  \end{equation}
  
Finally, the WLLS estimation is given as
  \begin{equation}\label{equ:24}
  {{\boldsymbol{\hat \beta }}_{:,{m_i}}} = {\left( {{\bf{C'}}} \right)^ + }{{\boldsymbol{\alpha }}_i'}.
  \end{equation}

\textit{Remark 5:} By substituting \eqref{equ:21} into \eqref{equ:24}, it can be seen that the estimation of \eqref{equ:24} has a quasi-static phase bias of
  \begin{equation}\label{equ:25}
    {\left( {{\bf{C'}}} \right)^ + }\left( {\left[ {\begin{array}{*{20}{c}}
    {\left| {{{{\bf{\hat h}}}_{:,1}}} \right|}\\
     \vdots \\
    {\left| {{{{\bf{\hat h}}}_{:,{N_t}}}} \right|}
    \end{array}} \right] \odot \left[ {\begin{array}{*{20}{c}}
    {{{\boldsymbol{\theta }}_{:,1}}}\\
     \vdots \\
    {{{\boldsymbol{\theta }}_{:,{N_t}}}}
    \end{array}} \right]} \right).
  \end{equation}
However, it will be shown in Sec.~\ref{sec:ea.ce} that the channel estimation algorithm can perfectly compensate for this bias, and result in a feasible estimation of \({{\bf{\Phi }}_{{m}}}{\bf{H}}{{\bf{\Psi }}_{{m}}}\), which is what we are really interested in rather than any other intermediate variables. Therefore, the bias in this step is not an issue.

\subsection{Step 3: Wiener Phase Estimation}
In this section, the inter-pilot-group phase sequence, which can be denoted as \(\left[ {{{{\boldsymbol{\hat \beta }}}_{:,{m_1}}}, \cdots ,{{{\boldsymbol{\hat \beta }}}_{:,{m_{{N_c}}}}}} \right]\), is considered. And an element-wise inter-pilot-group Wiener phase estimator is proposed to further suppress the phase estimation error.

By substituting \eqref{equ:21} into \eqref{equ:24}, the noise term of \({{\boldsymbol{\hat \beta }}_{:,{m_i}}}\) can be written as
  \begin{equation}\label{equ:26}
  \begin{aligned}
    {{\bf{n}}_{\hat \beta \left( {:,{m_i}} \right)}} = & {\left( {{\bf{C'}}} \right)^ + }{\left[ {{{\left( {{{{\bf{n}}}_{i,:,1}''}} \right)}^T}, \cdots ,{{\left( {{{{\bf{n}}}_{i,:,{N_t}}''}} \right)}^T}} \right]^T}\\
     = & \left[ {\begin{array}{*{20}{c}}
    {{{\boldsymbol{\varsigma }}_{1,:}}}\\
     \vdots \\
    {{{\boldsymbol{\varsigma }}_{{N_r} + {N_t} - 1,:}}}
    \end{array}} \right]\left[ {\begin{array}{*{20}{c}}
    {{{{\bf{n}}}_{i,:,1}''}}\\
     \vdots \\
    {{{{\bf{n}}}_{i,:,{N_t}}''}}
    \end{array}} \right] ,
  \end{aligned}
  \end{equation}
where \({{{\boldsymbol{\varsigma }}_{q,:}}}\) is the \(q^{th}\) row of \({\left( {{\bf{C'}}} \right)^ + }\).

Note the fact that \({n}_{i,k,l}'' \sim {\cal N}\left( {0,{{\sigma _n^2} \mathord{\left/ {\vphantom {{\sigma _n^2} {2{N_t}}}} \right. \kern-\nulldelimiterspace} {2{N_t}}}} \right)\) are i.i.d. AWGN variables, it is easy to obtain that \({n_{\hat \beta \left( {q,{m_i}} \right)}} \sim {\cal N}\left( {0,\sigma _{{n_W}\left( q \right)}^2} \right)\), where
  \begin{equation}\label{equ:27}
    \sigma _{{n_W}\left( q \right)}^2 = \frac{{{{\left\| {{{\boldsymbol{\varsigma }}_{q,:}}} \right\|}^2}}}{{2{N_t}}}\sigma _n^2 .
  \end{equation}
On the other hand, as a direct result from \eqref{equ:19}, the equivalent phase noise variance between adjacent pilot groups (e.g. \({{{ \beta }_{q,{m_i}}}}\) and \({{{ \beta }_{q,{m_{i+1}}}}}\)) can be given as
  \begin{equation}\label{equ:28}
    \sigma _{{p_W}\left( q \right)}^2 = \left\{ {\begin{aligned}
    & {L_c} \cdot \left( {\sigma _{\Delta \varphi }^2 + \sigma _{\Delta \psi }^2} \right) & ,\left( {q \le {N_r}} \right) &\\
    & {L_c} \cdot 2\sigma _{\Delta \psi }^2 & ,\left( {q > {N_r}} \right) & ,
    \end{aligned}} \right.
  \end{equation}
where \(\sigma _{\Delta \varphi }^2\) and \(\sigma _{\Delta \psi }^2\) are the variance of the phase innovations in \eqref{equ:3} and \eqref{equ:4}, respectively.

For the \(q^{th}\) row of the phase sequence, a balanced two-sided Wiener phase estimator with tap length \(L_{tap} = 2 \times {L_W} + 1\) can be given by
  \begin{equation}\label{equ:29}
    {\hat \beta _{q,{m_i}}}' = \sum\limits_{t =  - {L_W}}^{{L_W}} {{\omega _{t,q}}{{\hat \beta }_{q,{m_{i - t}}}}} ,
  \end{equation}
where \({{\boldsymbol{\omega }}_{:,q}} = {\left[ {{\omega _{ - {L_W},q}}, \cdots ,{\omega _{{L_W},q}}} \right]^T}\) are the corresponding Wiener coefficients. 

According to \cite{Ip2007p2675}, the coefficient vector can be calculated by
  \begin{equation}\label{equ:30}
  {{\mathbf{\omega }}_{:,q}} = \left( {{{\mathbf{K}}^{ - 1}}{{\mathbf{1}}_{\left( {2{L_W} + 1} \right) \times 1}}} \right){\left( {{{\mathbf{1}}_{1 \times \left( {2{L_W} + 1} \right)}}{{\mathbf{K}}^{ - 1}}{{\mathbf{1}}_{\left( {2{L_W} + 1} \right) \times 1}}} \right)^{ - 1}},
  \end{equation}
where the \(\left({2{L_W} + 1}\right) \times \left({2{L_W} + 1}\right)\) matrix \(\bf{K}\) is
  \begin{equation}\label{equ:31}
    {\bf{K}} = {{\bf{K}}_p} + {{\bf{K}}_n},
  \end{equation}
the element at the \(k^{th}\) row and \(l^{th}\) column of \({\bf{K}}_p\) is
  \begin{equation}\label{equ:32}
    {k_{p\left( {k,l} \right)}} = \left\{ {\begin{array}{*{20}{l}}
    {\sigma _{{p_W}\left( q \right)}^2 \cdot \max \left( {k,l} \right),}&{\left( {\max \left( {k,l} \right) \leqslant {L_W}} \right)} \\ 
    {\sigma _{{p_W}\left( q \right)}^2 \cdot \min \left( {k,l} \right),}&{\left( {\min \left( {k,l} \right) \geqslant {L_W} + 2} \right)} \\ 
    {0,}&{otherwise,} 
    \end{array}} \right.
  \end{equation}
and
  \begin{equation}\label{equ:33}
    {{\bf{K}}_n} = \sigma _{{n_W}\left( q \right)}^2 \cdot {{\bf{I}}_{\left( {2{L_W} + 1} \right)}}.
  \end{equation}
The terms \({\bf{K}}_p\) and \({\bf{K}}_n\), which optimize the coefficient vector, arise
from the autocorrelations of phase noise and AWGN, respectively.

\subsection{Step 4: Phase Recovery and Channel Estimation}\label{sec:ea.ce}
As the estimated phase given by \eqref{equ:29}, the full channel estimation (including phase) at the \(i^{th}\) pilot group can be calculated as
  \begin{equation}\label{equ:34}
    \begin{aligned}
    {{{\bf{\hat H}}}_i} = & \frac{1}{{{N_t}}}{\rm{diag}}\left( {{e^{ - j{{\hat \beta }_{1,{m_i}}}'}}, \cdots ,{e^{ - j{{\hat \beta }_{{N_r},{m_i}}}'}}} \right){{\bf{Y}}_i}{\bf{S}}_i^H\\
    & \times {\rm{diag}}\left( {{e^{ - j{{\hat \beta }_{{N_r} + 1,{m_i}}}'}}, \cdots ,{e^{ - j{{\hat \beta }_{{N_r} + {N_t} - 1,{m_i}}}'}},1} \right).
    \end{aligned}
  \end{equation}

And the channel estimation averaged over the whole frame can be given as
  \begin{equation}\label{equ:35}
    {\bf{\hat H}} = \frac{1}{{{N_c}}}\sum\limits_{i = 1}^{{N_c}} {{{{\bf{\hat H}}}_i}}.
  \end{equation}
  
Moreover, the phase at the \(m^{th}\) (\(m_i < m < m_{i+1}\)) symbol period can be approximated by linear interpolation as
  \begin{equation}\label{equ:36}
    {\hat \beta _{q,m}}' = {\hat \beta _{q,{m_i}}}' + \frac{{{{\hat \beta }_{q,{m_{i + 1}}}}' - {{\hat \beta }_{q,{m_i}}}'}}{{{m_{i + 1}} - {m_i}}}\left( {m - {m_i}} \right) .
  \end{equation}
  
Finally, the overall joint phase and channel estimation at the \(m^{th}\) symbol period can be calculated as
  \begin{equation}\label{equ:37}
    \begin{aligned}
    {{{\bf{\hat \Phi }}}_m}{\bf{\hat H}}{{{\bf{\hat \Psi }}}_m} = & {\rm{diag}}\left( {{e^{j{{\hat \beta }_{1,m}}'}}, \cdots ,{e^{j{{\hat \beta }_{{N_r},m}}'}}} \right){\bf{\hat H}}\\
    \times &  {\rm{diag}}\left( {{e^{j{{\hat \beta }_{{N_r} + 1,m}}'}}, \cdots ,{e^{j{{\hat \beta }_{{N_r} + {N_t} - 1,m}}'}},1} \right) .
    \end{aligned}
  \end{equation}


\textit{Remark 6:} As discussed earlier, the phase estimation of  \eqref{equ:24} has a quasi-static bias of \eqref{equ:25}. This bias also exists (but in the negative direction) in the elements of the diagonal matrices in \eqref{equ:34}. Therefore, the positive and negative bias is cancelled when calculating \eqref{equ:37}, which guarantees a feasible estimation of \({{{\bf{\hat \Phi }}}_m}{\bf{\hat H}}{{{\bf{\hat \Psi }}}_m}\).



\textit{Remark 7:} Online version of \eqref{equ:35} is preferable in real-time systems to reduce latency. And the \({\mathbf{\hat H}}\) at the \(i^{th}\) pilot group can be calculated as
  \begin{equation}\label{equ:r2}
    {{\mathbf{\hat H}}_{\left( i \right)}} = \left( {1 - K} \right){{\mathbf{\hat H}}_{\left( {i - 1} \right)}} + K{{\mathbf{\hat H}}_i},
  \end{equation}
where \(K\) is the updating factor which should be less than the coherent time of the channel.

\subsection{Computational Complexity Analysis}
In this paper, the computational complexity is defined as the number of complex addition (\(C{C^{\left[ A \right]}}\)) and multiplication (\(C{C^{\left[ M \right]}}\)) required to calculate the phase and channel information at each symbol.
Moreover, the computational complexity of one complex division is defined as one complex multiplication due to their similar computational complexity.
Furthermore, we assume that the complex exponential functions and the square root operations are implemented by means of a lookup table.

The channel amplitude estimation is calculated by \eqref{equ:6}, \eqref{equ:9}, and \eqref{equ:10}. 
Therefore, the computational complexity for a symbol can be calculated by
  \begin{equation}\label{equ:cc.01a}
    CC_{\left| {{\mathbf{\hat H}}} \right|}^{\left[ {\text{M}} \right]} = \underbrace {{{{N_r}N_t^2} \mathord{\left/
 {\vphantom {{{N_r}N_t^2} {{L_c}}}} \right.
 \kern-\nulldelimiterspace} {{L_c}}}}_{\eqref{equ:6}} + \underbrace {{{{N_r}{N_t}} \mathord{\left/
 {\vphantom {{{N_r}{N_t}} {{L_c}}}} \right.
 \kern-\nulldelimiterspace} {{L_c}}}}_{\eqref{equ:9}},
  \end{equation}
  \begin{equation}\label{equ:cc.01b}
    CC_{\left| {{\mathbf{\hat H}}} \right|}^{\left[ {\text{A}} \right]} = \underbrace {{{{N_r}{N_t}\left( {{N_t} - 1} \right)} \mathord{\left/
 {\vphantom {{{N_r}{N_t}\left( {{N_t} - 1} \right)} {{L_c}}}} \right.
 \kern-\nulldelimiterspace} {{L_c}}}}_{\eqref{equ:6}} + \underbrace {{{{N_r}{N_t}} \mathord{\left/
 {\vphantom {{{N_r}{N_t}} {{L_c}}}} \right.
 \kern-\nulldelimiterspace} {{L_c}}} + {{{N_r}{N_t}} \mathord{\left/
 {\vphantom {{{N_r}{N_t}} {{L_f}}}} \right.
 \kern-\nulldelimiterspace} {{L_f}}}}_{\eqref{equ:9}}.
  \end{equation}

The WLLS one-shot estimation is calculated by \eqref{equ:13}, \eqref{equ:14}, \eqref{equ:22}-\eqref{equ:24}. However, \eqref{equ:13} can use the intermediate result of \({{\mathbf{Y}}_i}{\mathbf{S}}_i^H\) in \eqref{equ:6}. Therefore, the computational complexity for a symbol has the form of
  \begin{equation}\label{equ:cc.02a}
    \begin{aligned}
  CC_{{\text{WLLS}}}^{\left[ {\text{M}} \right]} & = \underbrace {{{{N_r}{N_t}} \mathord{\left/
 {\vphantom {{{N_r}{N_t}} {{L_c}}}} \right.
 \kern-\nulldelimiterspace} {{L_c}}}}_{\eqref{equ:22}} + \underbrace {{{\left( {{N_r} + {N_t} - 1} \right){N_r}{N_t}} \mathord{\left/
 {\vphantom {{\left( {{N_r} + {N_t} - 1} \right){N_r}{N_t}} {{L_f}}}} \right.
 \kern-\nulldelimiterspace} {{L_f}}}}_{\eqref{equ:23}} \\ 
  & + \underbrace {{{{N_r}{N_t}{{\left( {{N_r} + {N_t} - 1} \right)}^2}} \mathord{\left/
 {\vphantom {{{N_r}{N_t}{{\left( {{N_r} + {N_t} - 1} \right)}^2}} {{L_f}}}} \right.
 \kern-\nulldelimiterspace} {{L_f}}}}_{\eqref{equ:24}} \\ 
  & + \underbrace {{{{N_r}{N_t}\left( {{N_r} + {N_t} - 1} \right)} \mathord{\left/
 {\vphantom {{{N_r}{N_t}\left( {{N_r} + {N_t} - 1} \right)} {{L_c}}}} \right.
 \kern-\nulldelimiterspace} {{L_c}}}}_{\eqref{equ:24}} , \\
 \end{aligned}
  \end{equation}
  \begin{equation}\label{equ:cc.02b}
    \begin{aligned}
  CC_{{\text{WLLS}}}^{\left[ {\text{A}} \right]} & = \underbrace {{{{N_r}{N_t}{{\left( {{N_r} + {N_t} - 1} \right)}^2}} \mathord{\left/
 {\vphantom {{{N_r}{N_t}{{\left( {{N_r} + {N_t} - 1} \right)}^2}} {{L_f}}}} \right.
 \kern-\nulldelimiterspace} {{L_f}}}}_{\eqref{equ:24}} \hfill \\
  & + \underbrace {{{\left( {{N_r}{N_t} - 1} \right)\left( {{N_r} + {N_t} - 1} \right)} \mathord{\left/
 {\vphantom {{\left( {{N_r}{N_t} - 1} \right)\left( {{N_r} + {N_t} - 1} \right)} {{L_c}}}} \right.
 \kern-\nulldelimiterspace} {{L_c}}}}_{\eqref{equ:24}} \hfill .\\ 
    \end{aligned}
  \end{equation}

The Wiener estimation is calculated by \eqref{equ:27}-\eqref{equ:33}. However, \eqref{equ:28} and \eqref{equ:33} are constants. Therefore, the computational complexity for a symbol is
  \begin{equation}\label{equ:cc.03a}
  \begin{aligned}
  CC_{{\text{Wiener}}}^{\left[ {\text{M}} \right]} & = \underbrace {{{\left( {{N_r} + {N_t} - 1} \right){N_r}{N_t}} \mathord{\left/
 {\vphantom {{\left( {{N_r} + {N_t} - 1} \right){N_r}{N_t}} {{L_f}}}} \right.
 \kern-\nulldelimiterspace} {{L_f}}}}_{\eqref{equ:27}} \hfill \\
  & + \underbrace {{{\left( {2{L_W} + 1} \right)\left( {{N_r} + {N_t} - 1} \right)} \mathord{\left/
 {\vphantom {{\left( {2{L_W} + 1} \right)\left( {{N_r} + {N_t} - 1} \right)} {{L_c}}}} \right.
 \kern-\nulldelimiterspace} {{L_c}}}}_{\eqref{equ:29}} \hfill \\
  & + \underbrace {{{\left[ {{{\left( {2{L_W} + 1} \right)}^3} + \left( {2{L_W} + 1} \right)} \right]} \mathord{\left/
 {\vphantom {{\left[ {{{\left( {2{L_W} + 1} \right)}^3} + \left( {2{L_W} + 1} \right)} \right]} {{L_f}}}} \right.
 \kern-\nulldelimiterspace} {{L_f}}}}_{\eqref{equ:30}} + \underbrace {{{2{L_W}} \mathord{\left/
 {\vphantom {{2{L_W}} {{L_f}}}} \right.
 \kern-\nulldelimiterspace} {{L_f}}}}_{\eqref{equ:32}} \hfill , \\
  \end{aligned}
  \end{equation}
  \begin{equation}\label{equ:cc.03b}
  \begin{aligned}
  CC_{{\text{Wiener}}}^{\left[ {\text{A}} \right]} & = \underbrace {{{\left( {{N_r} + {N_t} - 1} \right)\left( {{N_r}{N_t} - 1} \right)} \mathord{\left/
 {\vphantom {{\left( {{N_r} + {N_t} - 1} \right)\left( {{N_r}{N_t} - 1} \right)} {{L_f}}}} \right.
 \kern-\nulldelimiterspace} {{L_f}}}}_{\eqref{equ:27}} \hfill \\
  & + \underbrace {{{2{L_W}\left( {{N_r} + {N_t} - 1} \right)} \mathord{\left/
 {\vphantom {{2{L_W}\left( {{N_r} + {N_t} - 1} \right)} {{L_c}}}} \right.
 \kern-\nulldelimiterspace} {{L_c}}}}_{\eqref{equ:29}} \hfill \\
  & + \underbrace {{{\left[ {\left( {2{L_W} + 1} \right) \cdot 2{L_W} + 2{L_W}} \right]} \mathord{\left/
 {\vphantom {{\left[ {\left( {2{L_W} + 1} \right) \cdot 2{L_W} + 2{L_W}} \right]} {{L_f}}}} \right.
 \kern-\nulldelimiterspace} {{L_f}}}}_{\eqref{equ:30}} + \underbrace {{{2{L_W}} \mathord{\left/
 {\vphantom {{2{L_W}} {{L_f}}}} \right.
 \kern-\nulldelimiterspace} {{L_f}}}}_{\eqref{equ:31}} \hfill . \\ 
  \end{aligned}
  \end{equation}

The phase recovery and channel estimation is calculated by \eqref{equ:34}-\eqref{equ:37}. Again, \({{\mathbf{Y}}_i}{\mathbf{S}}_i^H\) in \eqref{equ:34} can use the intermediate result of \eqref{equ:6}. Therefore, the computational complexity for a symbol can be represented as
  \begin{equation}\label{equ:cc.04a}
  \begin{aligned}
  CC_{{\mathbf{\hat H}}}^{\left[ {\text{M}} \right]} & = \underbrace {{{\left( {2{N_r}{N_t} - {N_r}} \right)} \mathord{\left/
 {\vphantom {{\left( {2{N_r}{N_t} - {N_r}} \right)} {{L_c}}}} \right.
 \kern-\nulldelimiterspace} {{L_c}}}}_{\eqref{equ:34}} + \underbrace {\left( {{N_r} + {N_t} - 1} \right){{{L_d}} \mathord{\left/
 {\vphantom {{{L_d}} {{L_c}}}} \right.
 \kern-\nulldelimiterspace} {{L_c}}}}_{\eqref{equ:36}} \hfill \\
  & + \underbrace {\left( {2{N_r}{N_t} - {N_r}} \right){{{L_d}} \mathord{\left/
 {\vphantom {{{L_d}} {{L_c}}}} \right.
 \kern-\nulldelimiterspace} {{L_c}}}}_{\eqref{equ:37}} \hfill ,\\ 
    \end{aligned}
  \end{equation}
  \begin{equation}\label{equ:cc.04b}
    CC_{{\mathbf{\hat H}}}^{\left[ {\text{A}} \right]} = \underbrace {{{{N_r}{N_t}} \mathord{\left/
 {\vphantom {{{N_r}{N_t}} {{L_c}}}} \right.
 \kern-\nulldelimiterspace} {{L_c}}}}_{\eqref{equ:35}} + \underbrace {{{\left( {{N_r} + {N_t} - 1} \right)\left( {1 + 2{L_d}} \right)} \mathord{\left/
 {\vphantom {{\left( {{N_r} + {N_t} - 1} \right)\left( {1 + 2{L_d}} \right)} {{L_c}}}} \right.
 \kern-\nulldelimiterspace} {{L_c}}}}_{\eqref{equ:36}}.
  \end{equation}

Combining \eqref{equ:cc.01a}-\eqref{equ:cc.04b}, the overall computational complexity for each symbol is
  \begin{equation}\label{equ:cc.05}
    \begin{aligned}
      CC = & {C_M}\left( {CC_{\left| {{\mathbf{\hat H}}} \right|}^{\left[ {\text{M}} \right]} + CC_{{\text{WLLS}}}^{\left[ {\text{M}} \right]} + CC_{{\text{Wiener}}}^{\left[ {\text{M}} \right]} + CC_{{\mathbf{\hat H}}}^{\left[ {\text{M}} \right]}} \right) \hfill \\
      & + \left( {CC_{\left| {{\mathbf{\hat H}}} \right|}^{\left[ {\text{A}} \right]} + CC_{{\text{WLLS}}}^{\left[ {\text{A}} \right]} + CC_{{\text{Wiener}}}^{\left[ {\text{A}} \right]} + CC_{{\mathbf{\hat H}}}^{\left[ {\text{A}} \right]}} \right) \hfill ,\\ 
    \end{aligned} 
  \end{equation}
where \(C_M\) is the weighing coefficient for multiplication, indicating that the multiplication is much more complex than addition.

\textit{Remark 8:} The computational complexity of different estimation algorithms is compared numerically in Table~\ref{tab:1}. The weighing coefficient for multiplication is set at \({C_M} = 1\) for a fair and compatible comparison with existing researches. The pilot rate is set at \({R_p} = 0.1\). The frame length is set at \(L_f = 10^5\). For the SPA-MAP algorithm in \cite{krishnan2015algorithms}, we set the modulation format as binary phase shift keying (BPSK), and the coefficients as \(L_{det}=10\), \(L_{dec}=1\), \(r_c=0.5\), \(w_c=3\), \(w_r=6\). As shown in Table~\ref{tab:1}, the proposed WLLS-Wiener algorithm has a much lower computational complexity compared to the existing algorithms \cite{mehrpouyan2012joint,nasir2013phase,krishnan2015algorithms}.

\begin{table}[htb]
  \small
  \caption{Computational Complexity of different algorithms}
  \label{tab:1}
    \begin{center} 
      \renewcommand{\arraystretch}{1.3}
      \begin{tabular}{|c|c|c|c|}
        \hline
        MIMO & \(2 \times 2\) & \(4 \times 4\) & \(8 \times 8\) \\
        \hline
        \(L_W = 5\) & 19.6 & 58.5 & 193.5\\
        \hline
        \(L_W = 50\) & 57.0 & 100.4 & 237.6 \\        \hline
        EKF in \cite{nasir2013phase} & 3.6e2 & 3.7e3 & 6.8e4 \\
        \hline
        EKF-EKS in \cite{nasir2013phase} & 4.8e2 & 5.1e3 & 8.1e4 \\
        \hline
        EKF in \cite{mehrpouyan2012joint} & 1.2e3 & 5.6e5 & 2.8e8 \\
        \hline
        SPA-MAP in \cite{krishnan2015algorithms} & 5.9e3 & 1.0e5 & 1.1e7 \\
        \hline
        Online MAP in \cite{nasir2013phase} & 5.1e6 & 7.8e7 & 1.1e9 \\
        \hline
        Offline MAP in \cite{nasir2013phase} & 1.0e8 & 1.5e9 & 2.2e10 \\
        \hline
      \end{tabular}
    \end{center}
\end{table}

\section{Cramér–Rao Lower Bound}
\label{sec:crlb}

\subsection{CRLB for One-shot Phase Estimation}
An important question is how accurate the phase of different transmit and receive oscillators can be estimated in one pilot group (which is defined in assumption (A1)). Without loss of generality, the \(1^{st}\) pilot group is considered in this subsection. And all the corresponding indices are restricted to the \(1^{st}\) pilot group (i.e. \(i = 1\), \(1  \le m \le {N_t}\), \({m_i} = {m_1} = \left\lceil {{N_t}/2} \right\rceil \), \({\bf{Y}} = {{\bf{Y}}_1}\), \({\bf{S}} = {{\bf{S}}_1}\), and \({\bf{N}} = {{\bf{N}}_1}\)).

As shown in Appendix~\ref{Appendix:A}, the information of \({\varphi _{k,m_i}}\) and \({\psi _{l,m_i}}\) is only included in the angular terms of \({\bf{Y}}{{\bf{S}}^H}\), which are also the observed data of the CRLB in this subsection.

For the \(1^{st}\) pilot group, \eqref{equ:2} can be rewritten as
  \begin{equation}\label{equ:crlb.01}
    \begin{aligned}
    {y_{k,m}} = & \sum\limits_{l = 1}^{{N_t}} {{h_{k,l}}{e^{j\left( {{\varphi _{k,m}} + {\psi _{l,m}}} \right)}}{s_{l,m}}}  + {n_{k,m}}\\
     = & \sum\limits_{l = 1}^{{N_t}} {{h_{k,l}}{e^{j\left( {{\varphi _{k,{m_i}}} + {\psi _{l,{m_i}}} + {\gamma _{k,l,m}}} \right)}}{s_{l,m}}}  + {n_{k,m}},
    \end{aligned}
  \end{equation}
where
  \begin{equation}\label{equ:crlb.02}
    \begin{aligned}
      {\gamma _{k,l,m}} = & \left( {{\varphi _{k,m}} - {\varphi _{k,{m_i}}}} \right) + \left( {{\psi _{l,m}} - {\psi _{l,{m_i}}}} \right) \hfill \\
       = & \left\{ {\begin{aligned}
        - & \sum\limits_{m' = m + 1}^{{m_i}} {\left( {\Delta {\varphi _{k,m'}} + \Delta {\psi _{l,m'}}} \right)} , & \left( {m < {m_i}} \right) & \\ 
      & \sum\limits_{m' = {m_i} + 1}^m {\left( {\Delta {\varphi _{k,m'}} + \Delta {\psi _{l,m'}}} \right)} , & \left( {m > {m_i}} \right) & \\ 
      & 0, & \left( {m = {m_i}} \right) & .
    \end{aligned}} \right. \hfill \\ 
    \end{aligned}
  \end{equation}

For practical oscillators, the phase noise innovations are small \cite{mcneill1997jitter, hajimiri1999jitter}, and the approximation below holds 
  \begin{equation}\label{equ:crlb.03}
    {e^{j{\gamma _{k,l,m}}}} \approx 1 + j{\gamma _{k,l,m}}.
  \end{equation}
Therefore, \eqref{equ:crlb.01} can be approximated by
  \begin{equation}\label{equ:crlb.04}
    \begin{aligned}
    {y_{k,m}} \approx & \sum\limits_{l = 1}^{{N_t}} {{h_{k,l}}{e^{j\left( {{\varphi _{k,{m_i}}} + {\psi _{l,{m_i}}}} \right)}}{s_{l,m}}} \\
    + & j\sum\limits_{l = 1}^{{N_t}} {{h_{k,l}}{e^{j\left( {{\varphi _{k,{m_i}}} + {\psi _{l,{m_i}}}} \right)}}{\gamma _{k,l,m}}{s_{l,m}}}  + {n_{k,m}}.
    \end{aligned}
  \end{equation}
And \eqref{equ:crlb.04} can be rewritten as
  \begin{equation}\label{equ:crlb.05}
    \begin{aligned}
    {{\bf{y}}_{k,:}} \approx & \sum\limits_{l = 1}^{{N_t}} {{h_{k,l}}{e^{j\left( {{\varphi _{k,{m_i}}} + {\psi _{l,{m_i}}}} \right)}}{{\bf{s}}_{l,:}}} \\
     + & j\sum\limits_{l = 1}^{{N_t}} {{h_{k,l}}{e^{j\left( {{\varphi _{k,{m_i}}} + {\psi _{l,{m_i}}}} \right)}}\left( {{{\bf{s}}_{l,:}} \odot {{\boldsymbol{\gamma }}_{k,l,:}}} \right)}  + {{\bf{n}}_{k,:}} .
    \end{aligned}
  \end{equation}
Note the fact that \({\bf{Y}}{{\bf{S}}^H}\) is an orthogonal transformation of the observed data \(\bf{Y}\), it does not change any information if the observed data is given by \({\bf{Y}}{{\bf{S}}^H}\). Moreover, as shown in Appendix~\ref{Appendix:A}, the information of \({\varphi _{k,m_i}}\) and \({\psi _{l,m_i}}\) is only included in the angular terms of \({\bf{Y}}{{\bf{S}}^H}\). When the perfect channel estimation is assumed, which is the optimal case of the phase estimation, the element at the \(k^{th}\) row and \(l^{th}\) column of the angular terms of the observed data matrix \(\bf{O}\) can be given by Appendix~\ref{Appendix:A} as
  \begin{equation}\label{equ:crlb.06}
    \begin{aligned}
    {o_{k,l}} = & {\varphi _{k,{m_i}}} + {\psi _{l,{m_i}}} + \frac{{n_{k,l}^{\left( 2 \right)}}}{{\left| {{h_{k,l}}} \right|}}\\
     & - \sum\limits_{m' = 2}^{{m_i}} {\left( {\sum\limits_{m = 1}^{m' - 1} {\sum\limits_{l' = 1}^{{N_t}} {{\xi _{k,l,l',m}}} } } \right)\Delta {\varphi _{k,m'}}} \\
     & - \sum\limits_{l' = 1}^{{N_t}} {\sum\limits_{m' = 2}^{{m_i}} {\left( {\sum\limits_{m = 1}^{m' - 1} {{\xi _{k,l,l',m}}} } \right)\Delta {\psi _{l',m'}}} } \\
     & + \sum\limits_{m' = {m_i} + 1}^{{N_t}} {\left( {\sum\limits_{m = m'}^{{N_t}} {\sum\limits_{l' = 1}^{{N_t}} {{\xi _{k,l,l',m}}} } } \right)\Delta {\varphi _{k,m'}}} \\
     & + \sum\limits_{l' = 1}^{{N_t}} {\sum\limits_{m' = {m_i} + 1}^{{N_t}} {\left( {\sum\limits_{m = m'}^{{N_t}} {{\xi _{k,l,l',m}}} } \right)\Delta {\psi _{l',m'}}} } ,
    \end{aligned}
  \end{equation}
where \(n_{k,l}^{\left( 2 \right)} \sim {\cal N}\left( {0,{{\sigma _n^2} \mathord{\left/
 {\vphantom {{\sigma _n^2} {2{N_t}}}} \right.
 \kern-\nulldelimiterspace} {2{N_t}}}} \right)\) is i.i.d. real AWGN, and
  \begin{equation}\label{equ:crlb.07}
    {\xi _{k,l,l',m}} \buildrel \Delta \over = \Re \left( {\frac{{{h_{k,l'}}}}{{{N_t}{h_{k,l}}}}{e^{j\left( {{\psi _{l',{m_i}}} - {\psi _{l,{m_i}}}} \right)}}{s_{l',m}}s_{l,m}^*} \right).
  \end{equation}
It is also worth noting that the last four phase noise terms in \eqref{equ:crlb.06} are from the symbols before and after the reference index \(m_i\), respectively.

In order to calculate the Fisher information matrix, it is necessary to rearrange the angular terms of the observed data to vector form as
  \begin{equation}\label{equ:crlb.08}
    {\boldsymbol{\upsilon }} = {\left[ {{{\bf{o}}_{1,:}},{{\bf{o}}_{2,:}}, \cdots ,{{\bf{o}}_{{N_r},:}}} \right]^T}.
  \end{equation}

The expected value of \({\boldsymbol{\upsilon }}\) is denoted by \({{\boldsymbol{\mu }}_{\boldsymbol{\upsilon }}}\). Note the fact that all the noise terms in \eqref{equ:crlb.06} are zero mean i.i.d. AWGN, the elements of \({{\boldsymbol{\mu }}_{\boldsymbol{\upsilon }}}\) can therefore be represented as
  \begin{equation}\label{equ:crlb.09}
    {\mu _{\upsilon \left( {{N_t}\left( {k - 1} \right) + l} \right)}} = {\varphi _{k,{m_i}}} + {\psi _{l,{m_i}}}.
  \end{equation}
Considering that the variables to be estimated are given by \eqref{equ:19}, \eqref{equ:crlb.09} can be rewritten as
  \begin{equation}\label{equ:crlb.10}
    {\mu _{\upsilon \left( {{N_t}\left( {k - 1} \right) + l} \right)}} = \left\{ {\begin{aligned}
    & {\beta _{k,{m_i}}} + {\beta _{{N_r} + l,{m_i}}}, & \left( {l \ne {N_t}} \right)\\
    & {\beta _{k,{m_i}}}, & \left( {l = {N_t}} \right) & .
    \end{aligned}} \right.
  \end{equation}
And the first order derivative of \({{\boldsymbol{\mu }}_{\boldsymbol{\upsilon }}}\) can be calculated as
  \begin{equation}\label{equ:crlb.11}
    \frac{{\partial {{\boldsymbol{\mu }}_{\boldsymbol{\upsilon }}}}}{{\partial {\beta _{q,{m_i}}}}} = \left\{ {\begin{aligned}
    & {{\left[ {\underbrace {0, \cdots ,0}_{\left( {q - 1} \right){N_t}},\underbrace {1, \cdots ,1}_{{N_t}},\underbrace {0, \cdots ,0}_{\left( {{N_r} - q} \right){N_t}}} \right]}^T}, &  \left( {q \le {N_r}} \right)\\
    & {{\left[ {\overbrace {\overbrace {\underbrace {0, \cdots ,0}_{q - {N_r} - 1},1,\underbrace {0, \cdots ,0}_{{N_r} + {N_t} - q}}^{1 \times {N_t}}, \cdots }^{{\rm{repeat }}\ {N_r}\ {\rm{ times}}}} \right]}^T}, &  \left( {q > {N_r}} \right) &.
    \end{aligned}} \right. 
  \end{equation}
  
On the other hand, the covariance matrix \({{\bf{\Sigma }}_{\boldsymbol{\upsilon }}}\) of \({\boldsymbol{\upsilon }}\) is defined as

  \begin{equation}\label{equ:crlb.12}
    {{\bf{\Sigma }}_{\boldsymbol{\upsilon }}} \buildrel \Delta \over = E\left[ {\left( {{\boldsymbol{\upsilon }} - {{\boldsymbol{\mu }}_{\boldsymbol{\upsilon }}}} \right){{\left( {{\boldsymbol{\upsilon }} - {{\boldsymbol{\mu }}_{\boldsymbol{\upsilon }}}} \right)}^T}} \right].
  \end{equation}

Moreover, as shown in Appendix~\ref{Appendix:B}, the element at the \(\left( {{N_t}\left( {{k_1} - 1} \right) + {l_1}} \right) ^ {th}\) row and \(\left( {{N_t}\left( {{k_2} - 1} \right) + {l_2}} \right) ^{th}\) column of the covariance matrix \({{\bf{\Sigma }}_{\boldsymbol{\upsilon }}}\) has the form of
  \begin{equation}\label{equ:crlb.13}
    \begin{aligned}
    \Sigma & _{\upsilon  \left( {{N_t}\left( {{k_1} - 1} \right) + {l_1},{N_t}\left( {{k_2} - 1} \right) + {l_2}} \right)}\\
     = & \frac{{\delta \left( {{k_1} - {k_2}} \right)\delta \left( {{l_1} - {l_2}} \right)}}{{2{N_t}\left| {{h_{{k_1},{l_1}}}} \right|\left| {{h_{{k_2},{l_2}}}} \right|}}\sigma _n^2\\
     & + \sum\limits_{m' = 2}^{\left\lceil {{{{N_t}} \mathord{\left/
     {\vphantom {{{N_t}} 2}} \right.
     \kern-\nulldelimiterspace} 2}} \right\rceil } {\left[ \left( {\sum\limits_{{m_1} = 1}^{m' - 1} {\sum\limits_{{l_1'} = 1}^{{N_t}} {{\xi _{{k_1},{l_1},{l_1'},{m_1}}}} } } \right) \right.} \\
     & \left. \times \left( {\sum\limits_{{m_2} = 1}^{m' - 1} {\sum\limits_{{l_2'} = 1}^{{N_t}} {{\xi _{{k_2},{l_2},{l_2'},{m_2}}}} } } \right)\sigma _{\Delta \varphi }^2\delta \left( {{k_1} - {k_2}} \right) \right]\\
     & + \sum\limits_{l' = 1}^{{N_t}} {\sum\limits_{m' = 2}^{\left\lceil {{{{N_t}} \mathord{\left/
     {\vphantom {{{N_t}} 2}} \right.
     \kern-\nulldelimiterspace} 2}} \right\rceil } {\left[ \left( {\sum\limits_{{m_1} = 1}^{m' - 1} {{\xi _{{k_1},{l_1},l',{m_1}}}} } \right)\right.} } \\
     & \left. \times \left( {\sum\limits_{{m_2} = 1}^{m' - 1} {{\xi _{{k_2},{l_2},l',{m_2}}}} } \right)\sigma _{\Delta \psi }^2 \right]\\
     & + \sum\limits_{m' = \left\lceil {{{{N_t}} \mathord{\left/
     {\vphantom {{{N_t}} 2}} \right.
     \kern-\nulldelimiterspace} 2}} \right\rceil  + 1}^{{N_t}} {\left[ \left( {\sum\limits_{{m_1} = m'}^{{N_t}} {\sum\limits_{{l_1'} = 1}^{{N_t}} {{\xi _{{k_1},{l_1},{l_1'},{m_1}}}} } } \right)\right.} \\
     & \left. \times \left( {\sum\limits_{{m_2} = m'}^{{N_t}} {\sum\limits_{{l_2'} = 1}^{{N_t}} {{\xi _{{k_2},{l_2},{l_2'},{m_2}}}} } } \right)\sigma _{\Delta \varphi }^2\delta \left( {{k_1} - {k_2}} \right)\right]\\
     & + \sum\limits_{l' = 1}^{{N_t}} {\sum\limits_{m' = \left\lceil {{{{N_t}} \mathord{\left/
     {\vphantom {{{N_t}} 2}} \right.
     \kern-\nulldelimiterspace} 2}} \right\rceil  + 1}^{{N_t}} {\left[ \left( {\sum\limits_{{m_1} = m'}^{{N_t}} {{\xi _{{k_1},{l_1},l',{m_1}}}} } \right) \right.} } \\
    & \left. \times \left( {\sum\limits_{{m_2} = m'}^{{N_t}} {{\xi _{{k_2},{l_2},l',{m_2}}}} } \right)\sigma _{\Delta \psi }^2 \right],
    \end{aligned}
  \end{equation}
where \(\delta \left(  \cdot  \right)\) is the unit sampling function which is defined as
  \begin{equation}\label{equ:crlb.14}
    \delta \left( x \right) \buildrel \Delta \over = \left\{ {\begin{aligned}
    {1,\left( {x = 0} \right)}\\
    {0,\left( {x \ne 0} \right)} & .
    \end{aligned}} \right.
  \end{equation}

And the first order derivative of the covariance matrix can be calculated directly from \eqref{equ:crlb.13} as 
  \begin{equation}\label{equ:crlb.16}
    \begin{aligned}
    & \frac{{\partial {\Sigma _{\upsilon \left( {{N_t}\left( {{k_1} - 1} \right) + {l_1},{N_t}\left( {{k_2} - 1} \right) + {l_2}} \right)}}}}{{\partial {\beta _{q,{m_i}}}}}\\
    & = \sum\limits_{m' = 2}^{\left\lceil {{{{N_t}} \mathord{\left/
     {\vphantom {{{N_t}} 2}} \right.
     \kern-\nulldelimiterspace} 2}} \right\rceil } {\left[\left( {\sum\limits_{{m_1} = 1}^{m' - 1} {\sum\limits_{{l_1'} = 1}^{{N_t}} {\frac{{\partial {\xi _{{k_1},{l_1},{l_1'},{m_1}}}}}{{\partial {\beta _{q,{m_i}}}}}} } } \right)\right.} \\
     & \left. \times \left( {\sum\limits_{{m_2} = 1}^{m' - 1} {\sum\limits_{{l_2'} = 1}^{{N_t}} {{\xi _{{k_2},{l_2},{l_2'},{m_2}}}} } } \right)\sigma _{\Delta \varphi }^2\delta \left( {{k_1} - {k_2}} \right)\right]\\
     & + \sum\limits_{m' = 2}^{\left\lceil {{{{N_t}} \mathord{\left/
     {\vphantom {{{N_t}} 2}} \right.
     \kern-\nulldelimiterspace} 2}} \right\rceil } {\left[\left( {\sum\limits_{{m_1} = 1}^{m' - 1} {\sum\limits_{{l_1'} = 1}^{{N_t}} {{\xi _{{k_1},{l_1},{l_1'},{m_1}}}} } } \right)\right.} \\
     & \left. \times \left( {\sum\limits_{{m_2} = 1}^{m' - 1} {\sum\limits_{{l_2'} = 1}^{{N_t}} {\frac{{\partial {\xi _{{k_2},{l_2},{l_2'},{m_2}}}}}{{\partial {\beta _{q,{m_i}}}}}} } } \right)\sigma _{\Delta \varphi }^2\delta \left( {{k_1} - {k_2}} \right)\right]\\
     & + \sum\limits_{l' = 1}^{{N_t}} {\sum\limits_{m' = 2}^{\left\lceil {{{{N_t}} \mathord{\left/
     {\vphantom {{{N_t}} 2}} \right.
     \kern-\nulldelimiterspace} 2}} \right\rceil } {\left[\left( {\sum\limits_{{m_1} = 1}^{m' - 1} {\frac{{\partial {\xi _{{k_1},{l_1},l',{m_1}}}}}{{\partial {\beta _{q,{m_i}}}}}} } \right)\right.} } \\
     & \left. \times \left( {\sum\limits_{{m_2} = 1}^{m' - 1} {{\xi _{{k_2},{l_2},l',{m_2}}}} } \right)\sigma _{\Delta \psi }^2\right]\\
     & + \sum\limits_{l' = 1}^{{N_t}} {\sum\limits_{m' = 2}^{\left\lceil {{{{N_t}} \mathord{\left/
     {\vphantom {{{N_t}} 2}} \right.
     \kern-\nulldelimiterspace} 2}} \right\rceil } {\left[\left( {\sum\limits_{{m_1} = 1}^{m' - 1} {{\xi _{{k_1},{l_1},l',{m_1}}}} } \right)\right.} } \\
     & \left. \times \left( {\sum\limits_{{m_2} = 1}^{m' - 1} {\frac{{\partial {\xi _{{k_2},{l_2},l',{m_2}}}}}{{\partial {\beta _{q,{m_i}}}}}} } \right)\sigma _{\Delta \psi }^2\right]\\
     & + \sum\limits_{m' = \left\lceil {{{{N_t}} \mathord{\left/
     {\vphantom {{{N_t}} 2}} \right.
     \kern-\nulldelimiterspace} 2}} \right\rceil  + 1}^{{N_t}} {\left[\left( {\sum\limits_{{m_1} = m'}^{{N_t}} {\sum\limits_{{l_1'} = 1}^{{N_t}} {\frac{{\partial {\xi _{{k_1},{l_1},{l_1'},{m_1}}}}}{{\partial {\beta _{q,{m_i}}}}}} } } \right)\right.} \\
     & \left. \times \left( {\sum\limits_{{m_2} = m'}^{{N_t}} {\sum\limits_{{l_2'} = 1}^{{N_t}} {{\xi _{{k_2},{l_2},{l_2'},{m_2}}}} } } \right)\sigma _{\Delta \varphi }^2\delta \left( {{k_1} - {k_2}} \right)\right]\\
     & + \sum\limits_{m' = \left\lceil {{{{N_t}} \mathord{\left/
     {\vphantom {{{N_t}} 2}} \right.
     \kern-\nulldelimiterspace} 2}} \right\rceil  + 1}^{{N_t}} {\left[\left( {\sum\limits_{{m_1} = m'}^{{N_t}} {\sum\limits_{{l_1'} = 1}^{{N_t}} {{\xi _{{k_1},{l_1},{l_1'},{m_1}}}} } } \right)\right.} \\
     & \left. \times \left( {\sum\limits_{{m_2} = m'}^{{N_t}} {\sum\limits_{{l_2'} = 1}^{{N_t}} {\frac{{\partial {\xi _{{k_2},{l_2},{l_2'},{m_2}}}}}{{\partial {\beta _{q,{m_i}}}}}} } } \right)\sigma _{\Delta \varphi }^2\delta \left( {{k_1} - {k_2}} \right)\right]\\
     & + \sum\limits_{l' = 1}^{{N_t}} {\sum\limits_{m' = \left\lceil {{{{N_t}} \mathord{\left/
     {\vphantom {{{N_t}} 2}} \right.
     \kern-\nulldelimiterspace} 2}} \right\rceil  + 1}^{{N_t}} {\left[\left( {\sum\limits_{{m_1} = m'}^{{N_t}} {\frac{{\partial {\xi _{{k_1},{l_1},l',{m_1}}}}}{{\partial {\beta _{q,{m_i}}}}}} } \right)\right.} } \\
     & \left. \times \left( {\sum\limits_{{m_2} = m'}^{{N_t}} {{\xi _{{k_2},{l_2},l',{m_2}}}} } \right)\sigma _{\Delta \psi }^2\right]\\
     & + \sum\limits_{l' = 1}^{{N_t}} {\sum\limits_{m' = \left\lceil {{{{N_t}} \mathord{\left/
     {\vphantom {{{N_t}} 2}} \right.
     \kern-\nulldelimiterspace} 2}} \right\rceil  + 1}^{{N_t}} {\left[\left( {\sum\limits_{{m_1} = m'}^{{N_t}} {{\xi _{{k_1},{l_1},l',{m_1}}}} } \right)\right.} } \\
     & \left. \times \left( {\sum\limits_{{m_2} = m'}^{{N_t}} {\frac{{\partial {\xi _{{k_2},{l_2},l',{m_2}}}}}{{\partial {\beta _{q,{m_i}}}}}} } \right)\sigma _{\Delta \psi }^2\right] ,
    \end{aligned}
  \end{equation}
where the first derivative of \eqref{equ:crlb.07}, which is calculated in Appendix~\ref{Appendix:C}, has the form of
  \begin{equation}\label{equ:crlb.15}
    \frac{{\partial {\xi _{k,l,l',m}}}}{{\partial {\beta _{q,{m_i}}}}} = \left\{ {\begin{aligned}
    \begin{aligned}
     - \Im \left( {\frac{{{h_{k,l'}}}}{{{N_t}{h_{k,l}}}}{e^{j\left( {{\psi _{l',{m_i}}} - {\psi _{l,{m_i}}}} \right)}}{s_{l',m}}s_{l,m}^*} \right),\\
    \left( {l \ne l' = q - {N_r}} \right)
    \end{aligned}\\
    \begin{aligned}
    \Im \left( {\frac{{{h_{k,l'}}}}{{{N_t}{h_{k,l}}}}{e^{j\left( {{\psi _{l',{m_i}}} - {\psi _{l,{m_i}}}} \right)}}{s_{l',m}}s_{l,m}^*} \right),\\
    \left( {l' \ne l = q - {N_r}} \right)
    \end{aligned}\\
    {0, \qquad\qquad\qquad\qquad\qquad \left( {otherwise} \right)} & .
    \end{aligned}} \right.
  \end{equation}

Using \eqref{equ:crlb.11}, \eqref{equ:crlb.13}, and \eqref{equ:crlb.16}, the element at the \({q_1}^{th}\) row and \({q_2}^{th}\) column of the Fisher information matrix \(\bf{FIM}\) can be represented by the classic form as \cite[p. 47, (3.31)]{kay1993fundamentals} 
  \begin{equation}\label{equ:crlb.17}
    {\text{FI}}{{\text{M}}_{{q_1},{q_2}}} = \frac{{\partial {\boldsymbol{\mu }}_{\boldsymbol{\upsilon }}^T}}{{\partial {\beta _{{q_1}}}}}{\mathbf{\Sigma }}_{\boldsymbol{\upsilon }}^{ - 1}\frac{{\partial {{\boldsymbol{\mu }}_{\boldsymbol{\upsilon }}}}}{{\partial {\beta _{{q_2}}}}} + \frac{1}{2}{\text{tr}}\left( {{\mathbf{\Sigma }}_{\boldsymbol{\upsilon }}^{ - 1}\frac{{\partial {{\mathbf{\Sigma }}_{\boldsymbol{\upsilon }}}}}{{\partial {\beta _{{q_1}}}}}{\mathbf{\Sigma }}_{\boldsymbol{\upsilon }}^{ - 1}\frac{{\partial {{\mathbf{\Sigma }}_{\boldsymbol{\upsilon }}}}}{{\partial {\beta _{{q_2}}}}}} \right).
  \end{equation}
And the CRLB for one-shot phase estimation can be calculated as \cite[p. 44, (3.24)]{kay1993fundamentals}
  \begin{equation}\label{equ:crlb.18}
  {\bf{CRLB}}\left( {\boldsymbol{\beta }} \right) = {\rm{diag}}\left( {{\bf{FI}}{{\bf{M}}^{ - 1}}} \right).
  \end{equation}

\textit{Remark 9:} The CRLB derived in this subsection is only dependent on the angular information of the observed data. This is different from previous researches such as \cite{mehrpouyan2012joint,nasir2013phase}. Considering that most known practical phase estimation algorithms only use phase information of the observed data for the phase estimation, this CRLB gives a more practical lower bound for the phase estimation in MIMO systems. Moreover, this CRLB can be easily modified into two simplified cases which can give intuitively satisfying explanations of the system. These two cases are discussed in Remark 10 and 11, respectively. And the numerical results of these two cases will be given in Sec.~\ref{sec:nr}.

\textit{Remark 10:} The first case can be simplified from \eqref{equ:crlb.06} as
  \begin{equation}\label{equ:crlb.19}
    {o_{k,l}} \approx {\varphi _{k,{m_i}}} + {\psi _{l,{m_i}}} + \frac{{n_{k,l}^{\left( 2 \right)}}}{{\left| {{h_{k,l}}} \right|}}.
  \end{equation}
This approximation is obtained from \eqref{equ:crlb.06} by neglecting all the phase noise terms. This describes the situation in the low SNR region where the AWGN is the dominant factor of noise. And it can be easily shown that the mean squared error (MSE) of phase estimation is proportional to \({\sigma _n^2}\).

\textit{Remark 11:} The second case can be simplified from \eqref{equ:crlb.06} as
  \begin{equation}\label{equ:crlb.20}
    \begin{aligned}
    {o_{k,l}} & \approx {\varphi _{k,{m_i}}} + {\psi _{l,{m_i}}}\\
     & - \frac{1}{{{N_t}}}\sum\limits_{m' = 2}^{{m_i}} {\sum\limits_{m = 1}^{m' - 1} {\Delta {\varphi _{k,m'}}} }  + \frac{1}{{{N_t}}}\sum\limits_{m' = {m_i} + 1}^{{N_t}} {\sum\limits_{m = m'}^{{N_t}} {\Delta {\varphi _{k,m'}}} } \\
     & - \frac{1}{{{N_t}}}\sum\limits_{m' = 2}^{{m_i}} {\sum\limits_{m = 1}^{m' - 1} {\Delta {\psi _{l,m'}}} }  + \frac{1}{{{N_t}}}\sum\limits_{m' = {m_i} + 1}^{{N_t}} {\sum\limits_{m = m'}^{{N_t}} {\Delta {\psi _{l,m'}}} } .
    \end{aligned}
  \end{equation}
This approximation is obtained from \eqref{equ:crlb.06} by setting \(l'=l\) and neglecting the AWGN term. According to \eqref{equ:crlb.07}, \({\xi _{k,l,l',m}} = 1/N_t\) when \(l'=l\). This describes the situation in the high SNR region when the system is only contaminated by a certain part of the phase noise. In this case, \({\varphi _{k,{m_i}}}\) is always degraded by corresponding \(\Delta {{\boldsymbol{\varphi }}_{k,:}}\). And \({\psi _{l,{m_i}}}\) is always degraded by corresponding \(\Delta {{\boldsymbol{\psi }}_{l,:}}\). Moreover, no cross term occurs in \eqref{equ:crlb.20}. Considering that the phase noise is a stationary process with independent increments, \eqref{equ:crlb.20} indicates that there is an MSE floor of the phase estimation in the high SNR region when we only exploit the angular terms of the observed data to estimate the phase information.

\subsection{CRLB for Element-wise Wiener Estimation}
Although the ultimate bound of the phase estimation performance in the MIMO system is very difficult to obtain, a very useful CRLB for element-wise Wiener estimation can be obtained to evaluate the performance of the proposed algorithm.

Note the fact that Wiener estimator is the optimal filter for the element-wise phase estimation problem \cite{Ip2007p2675,Proakis2007signal}, the CRLB is equal to the MSE performance of a two-sided infinite impulse response (IIR) Wiener filter.

Without loss of generality, we only consider the MSE performance of the Wiener estimation for the first pilot group. For the \(q^{th}\) IIR Wiener filter, \(\sigma _{pW\left( q \right)}^2\) is given by \eqref{equ:28}, \(\sigma _{\beta \left( q \right)}^2 = {\bf{CRLB}}\left( {{\beta _q}} \right)\) is given by \eqref{equ:crlb.18}, and the coefficient of the \(i^{th}\) pilot group can be rewritten from \eqref{equ:30} as \cite{Ip2007p2675}
  \begin{equation}\label{equ:crlb2.1}
    {\omega _{i,q}} = \left\{ {\begin{aligned}
    \frac{{\kappa \tau }}{{1 - {\kappa ^2}}}{\kappa ^{i - 1}}, & \left( {i \ge 1} \right)\\
    \frac{{\kappa \tau }}{{1 - {\kappa ^2}}}{\kappa ^{1 - i}}, & \left( {i < 1} \right) ,
    \end{aligned}} \right.
  \end{equation}
where
  \begin{equation}\label{equ:crlb2.2}
    \kappa  = \left( {1 + \tau /2} \right) - \sqrt {{{\left( {1 + \tau /2} \right)}^2} - 1},
  \end{equation}
  \begin{equation}\label{equ:crlb2.3}
    \tau  = \sigma _{pW\left( q \right)}^2/\sigma _{\beta \left( q \right)}^2.
  \end{equation}

By setting \({L_W} \to  + \infty \) in \eqref{equ:29}, and noting that the phase noise is a stationary process with independent increments, the lower bound of the MSE performance of the Wiener estimator can be written as
  \begin{equation}\label{equ:crlb2.4}
    \begin{aligned}
      \sigma _{W\left( q \right)}^2 = \sum\limits_{i =  - \infty }^{ + \infty } {\omega _{i,q}^2\sigma _{\beta \left( q \right)}^2} & + \sum\limits_{i = 2}^\infty  {{{\left( {\sum\limits_{i' = i}^\infty  {{\omega _{i',q}}} } \right)}^2}\sigma _{pW\left( q \right)}^2}  \hfill \\
      & + \sum\limits_{i =  - \infty }^0 {{{\left( {\sum\limits_{i' =  - \infty }^i {{\omega _{i',q}}} } \right)}^2}\sigma _{pW\left( q \right)}^2} . \hfill \\
    \end{aligned}
  \end{equation}
  
By substituting \eqref{equ:crlb2.1} into \eqref{equ:crlb2.4} and calculating the summation, the following equation holds
  \begin{equation}\label{equ:crlb2.5}
    \sigma _{W\left( q \right)}^2 = \sigma _{\beta \left( q \right)}^2\frac{{{\kappa ^2}{\tau ^2}\left( {1 + {\kappa ^2}} \right)}}{{{{\left( {1 + \kappa } \right)}^3}{{\left( {1 - \kappa } \right)}^3}}} + 2\sigma _{pW\left( q \right)}^2\frac{{{\kappa ^4}{\tau ^2}}}{{{{\left( {1 + \kappa } \right)}^3}{{\left( {1 - \kappa } \right)}^5}}}.
  \end{equation}
  
By substituting \eqref{equ:crlb2.2} and \eqref{equ:crlb2.3} into \eqref{equ:crlb2.5}, \(\sigma _W^2\) can be represented by \(\sigma _{\beta \left( q \right)}^2\) and \(\sigma _{pW\left( q \right)}^2\). Moreover, the equation can be simplified by Mathematica\(^\text{TM}\) 11.2.0.0 as
  \begin{equation}\label{equ:crlb2.6}
    \sigma _{W\left( q \right)}^2 = {\left[ {\frac{4}{{\sigma _{pW\left( q \right)}^2\sigma _{\beta \left( q \right)}^2}} + \frac{1}{{\sigma _{\beta \left( q \right)}^4}}} \right]^{ - \frac{1}{2}}}.
  \end{equation}

Wiener filter is the optimal filter for the element-wise phase estimation problem \cite{kay1993fundamentals}. Noting that the noise MSE in \eqref{equ:crlb2.6} is given by \eqref{equ:28} and \eqref{equ:crlb.18}, \eqref{equ:crlb2.6} is also the CRLB for the element-wise Wiener estimators.

\section{Numerical Results}
\label{sec:nr}
In this section, the numerical performance of our newly proposed phase and channel estimation algorithm is evaluated against the CRLB. Moreover, the BER performance of the proposed algorithm is also evaluated in details. It is assumed that  \(\sigma _{\Delta \varphi }^2 = \sigma _{\Delta \psi }^2 = \sigma _\Delta ^2\), and \(\sigma _n^2=1/{\rm{SNR}}\) throughout this section. Moreover, the elements of channel matrix \({\bf{H}}\) are assumed to be i.i.d. standard complex Gaussian distribution so that the amplitude of the matrix element is i.i.d. Rayleigh distributed variable. Furthermore, unless otherwise specified, below coefficients are used throughout this section: The transmission symbol format is quadrature phase shift keying (QPSK). The MIMO decoder uses the conventional minimum mean squared error (MMSE) algorithm in \cite{Proakis2008communications}. \(N_t=2\), \(N_r=2\), \(\sigma _\Delta ^2 = {10^{ - 4}}\), the pilot rate \({R_p}=1/10\), and the Wiener filter tap length \(L_{tap} = 101\). To evaluate the system performance, a minimum of \(10^5\) independent Monte-Carlo trials are used. And a frame length of \( L_f = 3 \times 10^3\) symbols are transmitted in each trial. The phase and channel estimation performance will be discussed in Sec.~\ref{sec:nr.phase}. And the BER performance will be discussed in \ref{sec:nr.ber}.

\subsection{Estimation Performance}
\label{sec:nr.phase}

  \begin{figure}[t]
  \centering
  \includegraphics[width=3.4in]{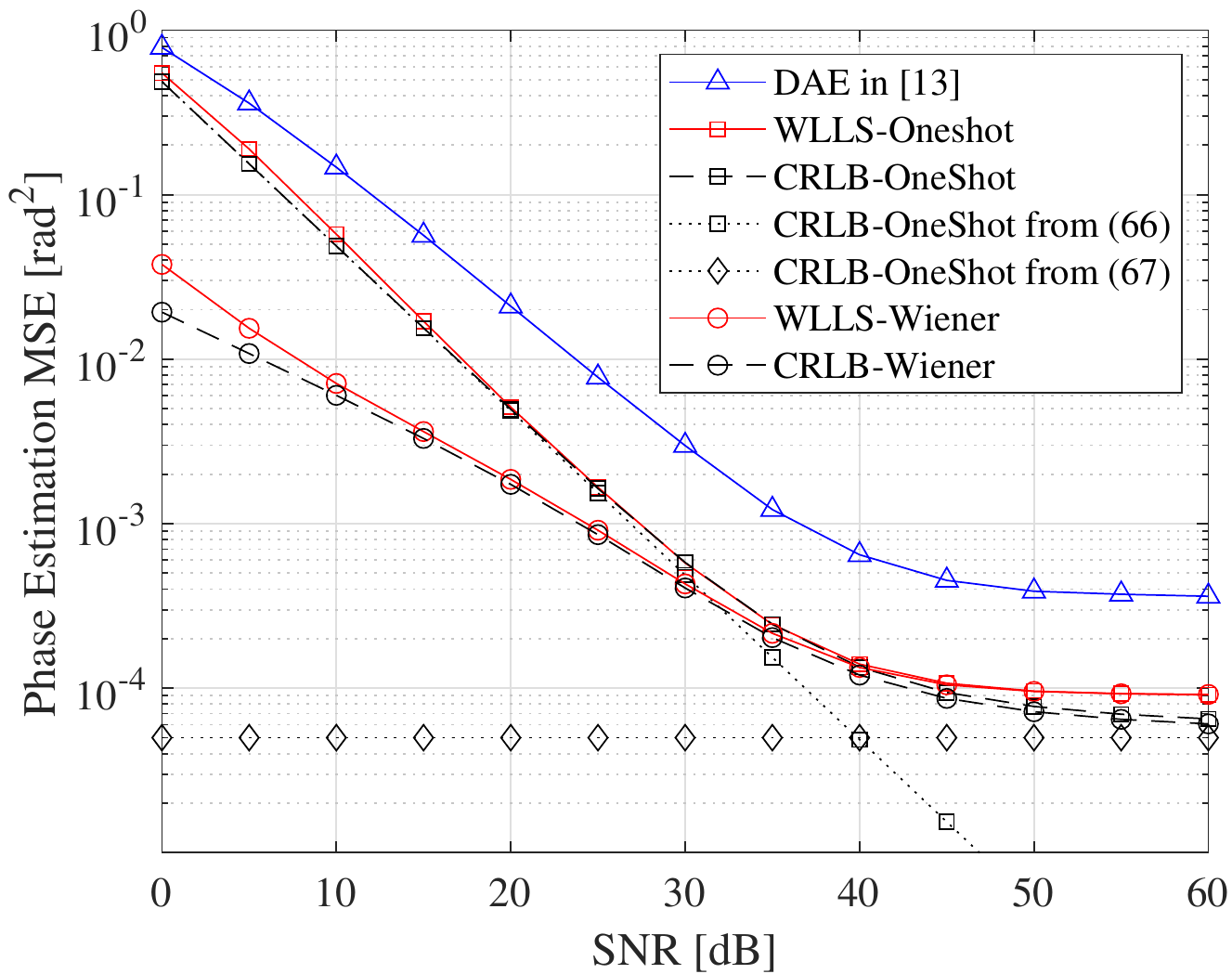}
  \caption{MSE of phase noise for a \(2 \times 2\) MIMO system. Different phase estimators. Lines - MSE performance of DAE in \cite{mehrpouyan2012joint} (triangles), our proposed algorithm without (squares) and with (circles) Wiener filtering. Dashed lines - CRLBs without (squares) and with (circles) Wiener filtering. Dotted lines - simplified CRLBs from \eqref{equ:crlb.19} (squares, neglecting phase noise), and \eqref{equ:crlb.20} (diamonds, neglecting AWGN and cross terms).}
  \label{fig:srPhaseMain}
  \end{figure}
The phase estimation MSE performance for both WLLS one-shot estimator and WLLS Wiener estimator are shown in Fig.~\ref{fig:srPhaseMain}. The results indicate that the proposed WLLS one-shot estimator performs better than the conventional data aided estimation (DAE) algorithm in \cite{mehrpouyan2012joint}. 
This is because the WLLS one-shot estimator extracts \(N_r + N_t -1 \) phase information from \(N_r N_t\) observed angular terms, and assigns a smaller weight to the observed data with larger errors, thereby reducing the phase estimation error (refer to \eqref{equ:21}-\eqref{equ:24} for details).
Moreover, the performance of the WLLS one-shot estimator is very close to the CRLB for one-shot estimation. 
The slight performance degradation in the low SNR region for the WLLS one-shot estimator is mainly due to the small noise assumption. The slight performance degradation in the high SNR region is mainly due to neglecting the intra-pilot-group phase noise.
On the other hand, introducing the Wiener estimator further improves the performance of the WLLS one-shot estimator in the low SNR region. This is because the Wiener filtering reduces the impact of AWGN. Moreover, the performance of the WLLS-Wiener estimator is very close to the CRLB for Wiener Estimation. The slight performance degradation in the low SNR region for the WLLS-Wiener estimator is mainly due to the small noise assumption and the finite filter tap length. The slight performance degradation in the high SNR region is mainly due to neglecting the intra-pilot-group phase noise.
Furthermore, the simplified CRLB from \eqref{equ:crlb.19} is a tight lower bound in the low SNR region, which indicates the AWGN dominant situation. And the simplified CRLB from \eqref{equ:crlb.20} indicates the MSE floor in the high SNR region, which is a loose lower bound because neglecting certain cross terms of the phase noise.

\begin{figure}[t]
  \centering
  \includegraphics[width=3.4in]{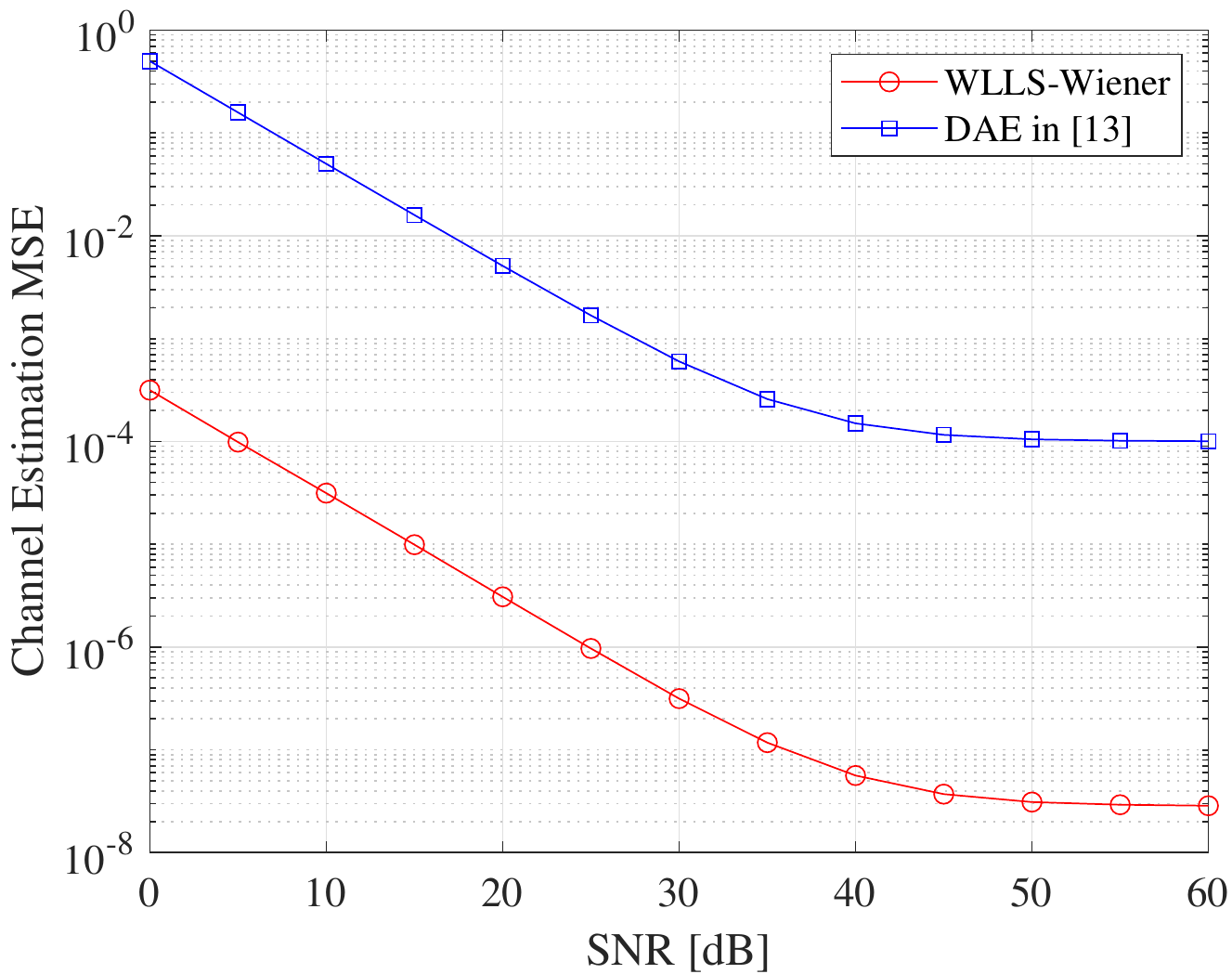}
  \caption{MSE of channel estimation for a \(2 \times 2\) MIMO system. A comparison between our proposed algorithm (circles) and DAE in \cite{mehrpouyan2012joint} (squares).}
  \label{fig:srChannelMain}
  \end{figure}
The channel estimation MSE is plotted in Fig.~\ref{fig:srChannelMain}. The results indicate that the proposed algorithm can obtain more than \(10^3\) times better channel estimation accuracy than the conventional DAE algorithm in \cite{mehrpouyan2012joint}. This is because the fast varying phase noise is cancelled before the full channel estimation in the proposed algorithm, and the channel estimation process can be averaged over the whole frame to improve the estimation accuracy. 

\begin{figure}[t]
  \centering
  \includegraphics[width=3.4in]{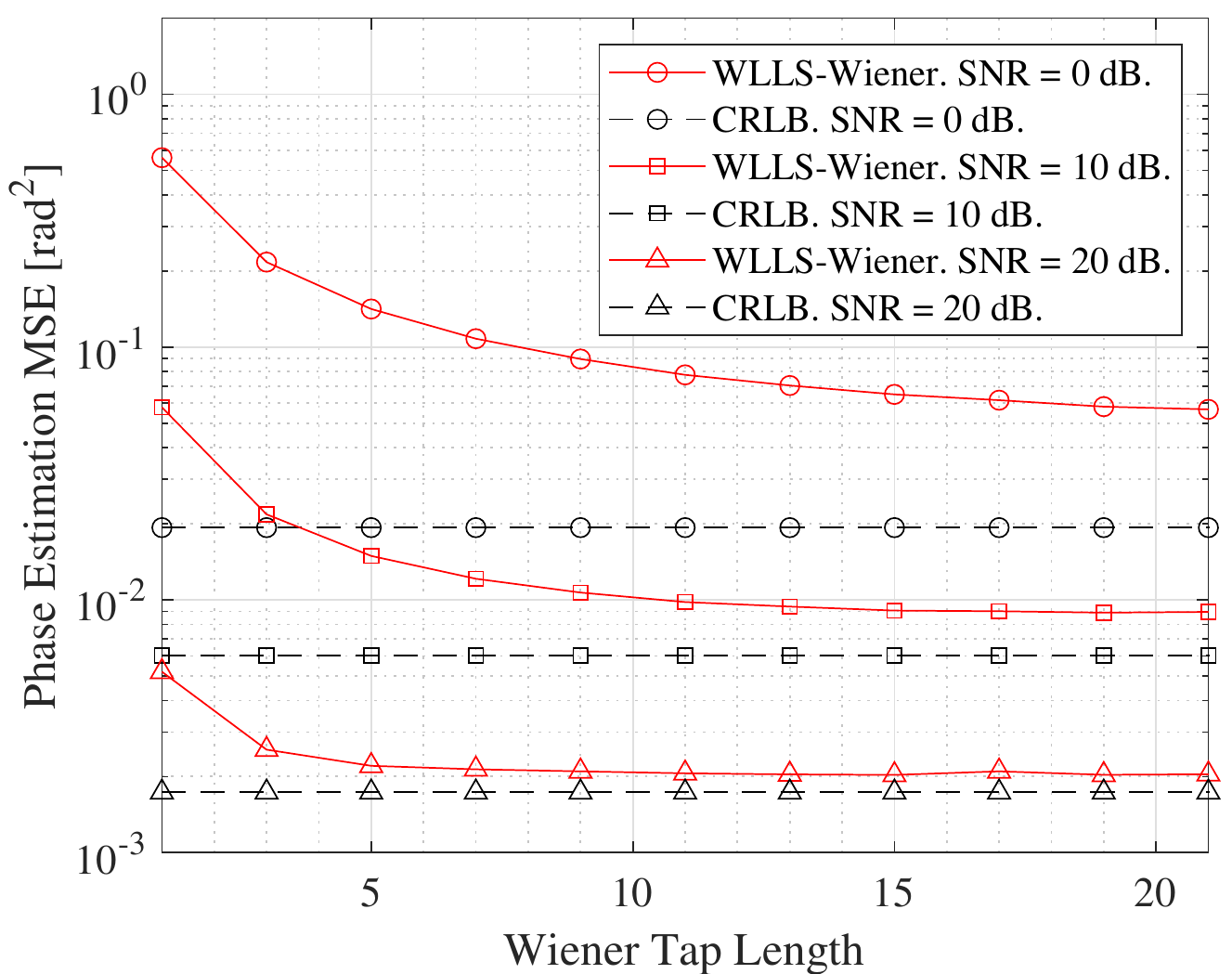}
  \caption{WLLS Wiener phase estimation for a \(2 \times 2\) MIMO system. MSE (red lines) and corresponding CRLB (black dashed lines). \(\rm{SNR}=0\) (circles), \(10\) (squares), \(20\) dB (triangles). Wiener filter tap length \(L_{tap}=1,3,5,...,21\).}
  \label{fig:srPhase2}
  \end{figure}

In order to quantify the influence of the Wiener filter tap length, Fig.~\ref{fig:srPhase2} compares the phase estimation MSE performance of WLLS Wiener phase estimator for different Wiener filter tap lengths (1, 3, 5,..., 21). As shown in Fig.~\ref{fig:srPhase2}, when the SNR is lower, the effective filter length will be larger and the curve converges slower. When the filter length is large and the curve converges, an MSE performance degradation is observed, and the performance degradation is larger when the SNR is lower. We believe this is mainly due to the small angle assumption in the derivation. When the filter tap length is small, we can observe more MSE performance degradation, which is due to the finite tap length effect.  Although finite tap length degrades the estimation performance and a larger tap length can give a better estimation accuracy, it is shown in Fig.~\ref{fig:srPhase2} that a relatively small tap length of 11 performs well when the phase noise variance \(\sigma_\Delta^2 = 10^{-4}\).

\begin{figure}[t]
  \centering
  \includegraphics[width=3.4in]{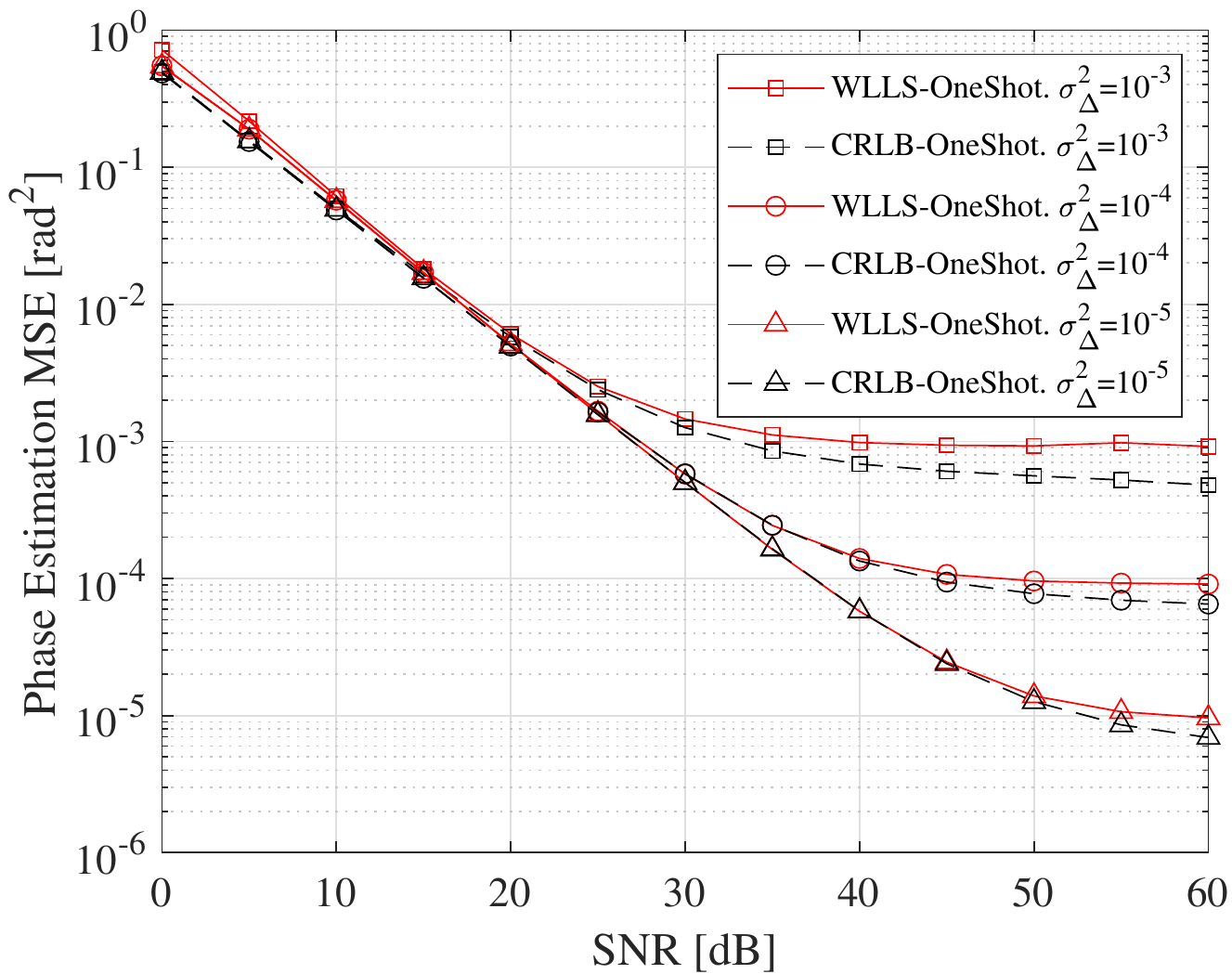}
  \caption{WLLS one-shot phase estimation for a \(2 \times 2\) MIMO system. MSE (red lines) and corresponding CRLB (black dashed lines). \(\sigma _\Delta ^2 =10^{-3}\) (squares), \(10^{-4}\) (circles), \(10^{-5}\) (triangles).}
  \label{fig:srPhase3}
  \end{figure}

Fig.~\ref{fig:srPhaseMain} indicates that the MSE has a minimum value, or floor in the High SNR region. In order to verify the origin of this floor, Fig.~\ref{fig:srPhase3} compares the phase estimation MSE performance of the WLLS one-shot phase estimator for different phase noise variances (\(\sigma _\Delta ^2 =10^{-3}, 10^{-4}, 10^{-5}\)). In the low SNR region, the MSE is similar as it is dominated by AWGN. In the high SNR region, the MSE floor is dominated by intra-pilot-group phase noise, which is proportional to the phase noise variance. Therefore, the MSE floor is also proportional to the phase noise variance. Moreover, similar to Fig.~\ref{fig:srPhaseMain}, the MSE penalty in the very high SNR region is due to neglecting the intra-pilot-group phase noise in the WLLS estimator.

\begin{figure}[t]
  \centering
  \includegraphics[width=3.4in]{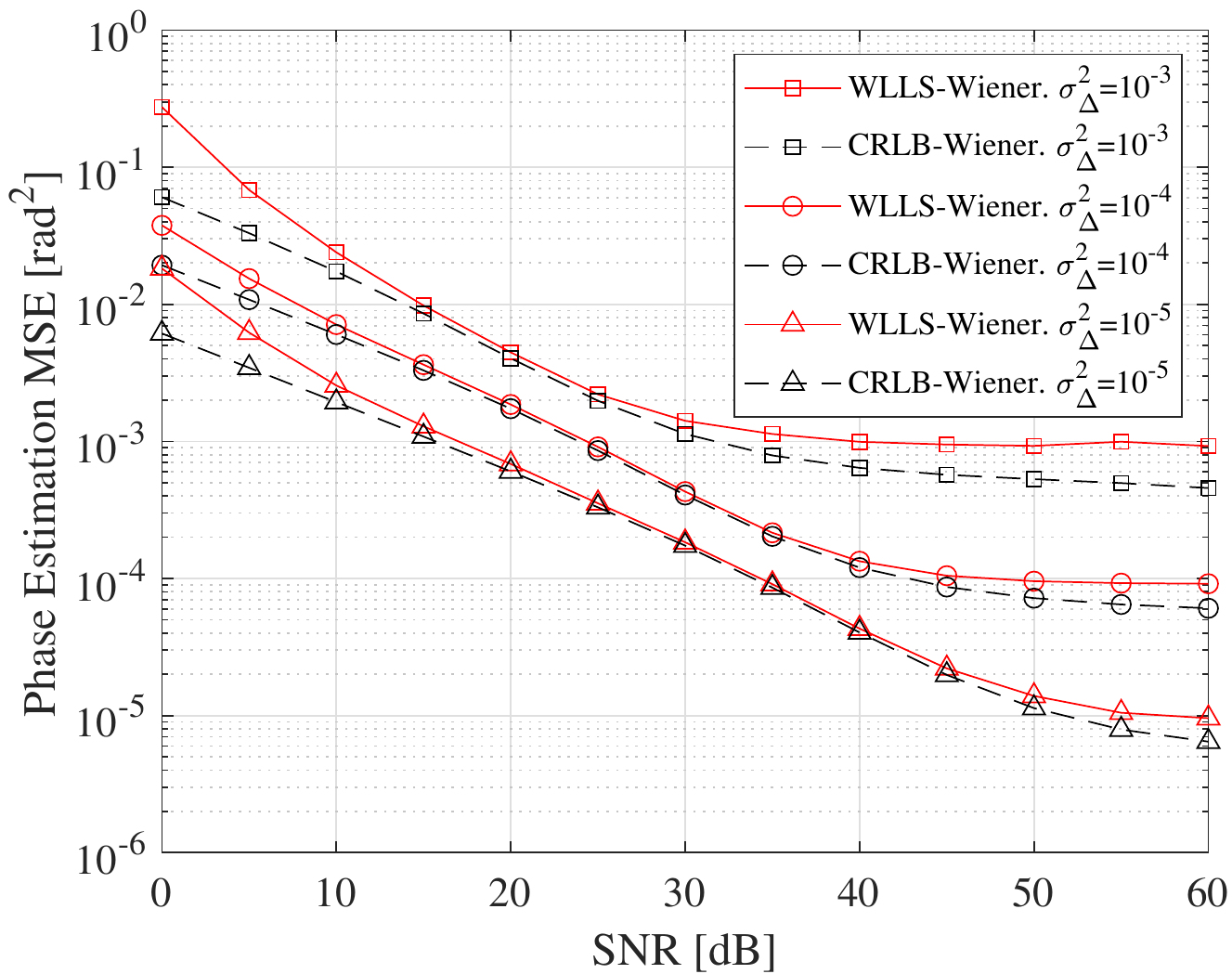}
  \caption{WLLS-Wiener phase estimation for a \(2 \times 2\) MIMO system. MSE (red line) and corresponding CRLB (black dashed line). \(\sigma _\Delta ^2 =10^{-3}\) (squares), \(10^{-4}\) (circles), \(10^{-5}\) (triangles).}
  \label{fig:srPhase4}
  \end{figure}

Fig.~\ref{fig:srPhase4} compares the phase estimation MSE performance of the WLLS Wiener phase estimator for different phase noise variances (\(\sigma _\Delta ^2 =10^{-3}, 10^{-4}, 10^{-5}\)). Unlike Fig.~\ref{fig:srPhase3}, where similar performance is observed for all phase noise variances in the low SNR region, a phase noise dependant MSE performance is observed. This is because the effective length of the Wiener filter is shorter when the phase noise is larger, and so there is less noise averaging.
Again, the MSE degradation in the low SNR region is mainly due to the small angle approximation.

\subsection{BER Performance of the MIMO system}
\label{sec:nr.ber}

  \begin{figure}[t]
  \centering
  \includegraphics[width=3.4in]{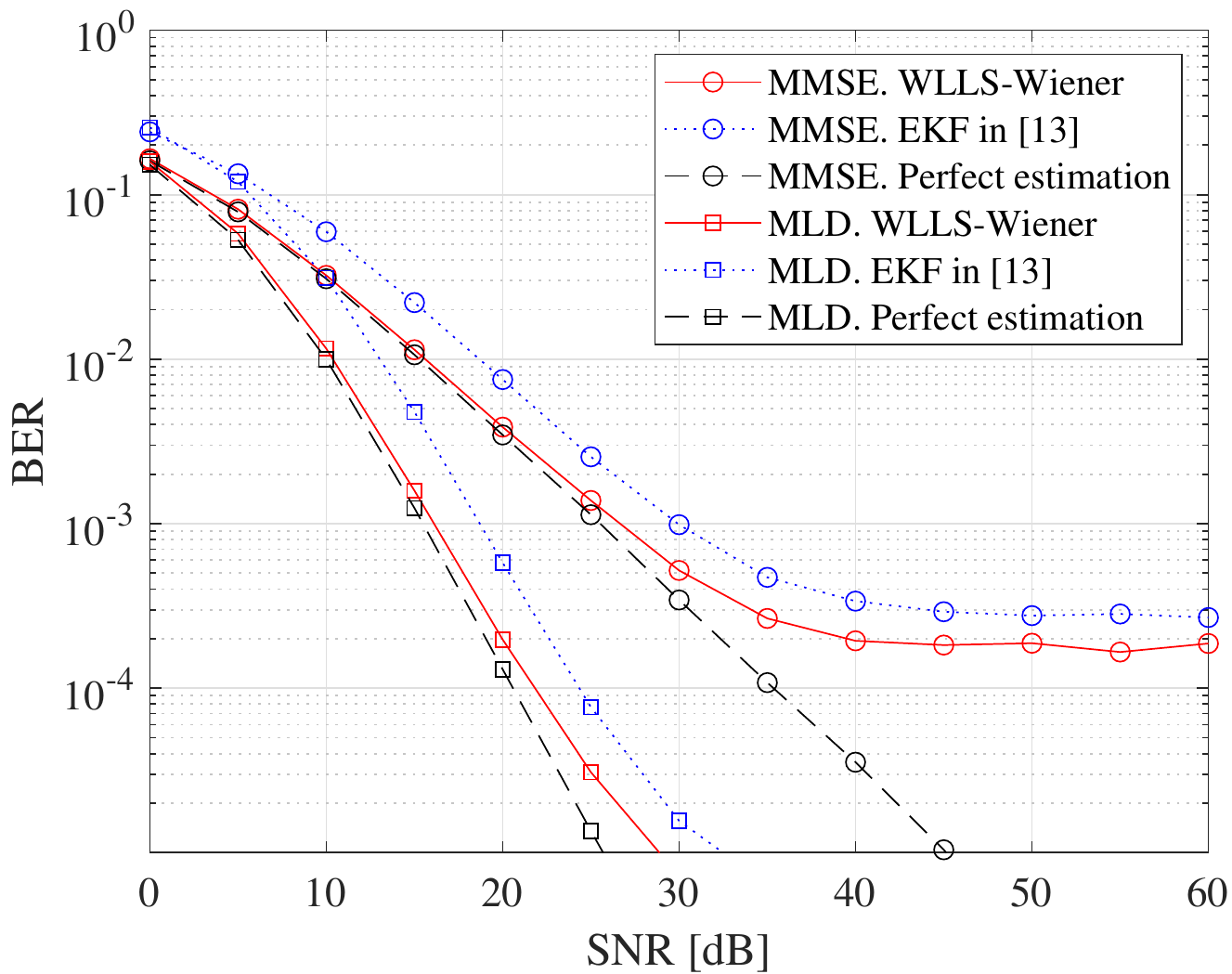}
  \caption{BER of a \(2 \times 2\) MIMO system with MMSE (circles) and MLD (squares) decoding. Red lines - proposed algorithm, blue dotted lines - EKF in \cite{mehrpouyan2012joint}, black dashed lines - perfect phase and channel estimation.}
  \label{fig:srBer1}
  \end{figure}
The BER performance of both maximum-likelihood decoder (MLD) and MMSE receiver are shown in Fig.~\ref{fig:srBer1}. Because of better phase and channel estimation accuracy, the proposed algorithm outperforms the conventional DAE algorithm. On the other hand, the BER floor is observed when using MMSE MIMO decoder, this is a direct consequence of the phase estimation MSE floor, which introduces an outage probability floor in the high SNR region. However, this problem can be suppressed by exploiting higher degree of freedom in MIMO systems (e.g. by applying MLD). As a result, when compared to the perfect phase and channel estimation scenario at the HD-FEC limit of \(\rm{BER} = 4.7 \times 10^{-3}\) \cite{alvarado2015replacing}, the proposed algorithm has an SNR penalty of approximately 0.5~dB for both MLD and MMSE MIMO decoders, while the conventional algorithm has an SNR penalty of approximately 3.2~dB and 3.5~dB for MLD and MMSE MIMO decoders, respectively. 

\begin{figure}[t]
  \centering
  \includegraphics[width=3.4in]{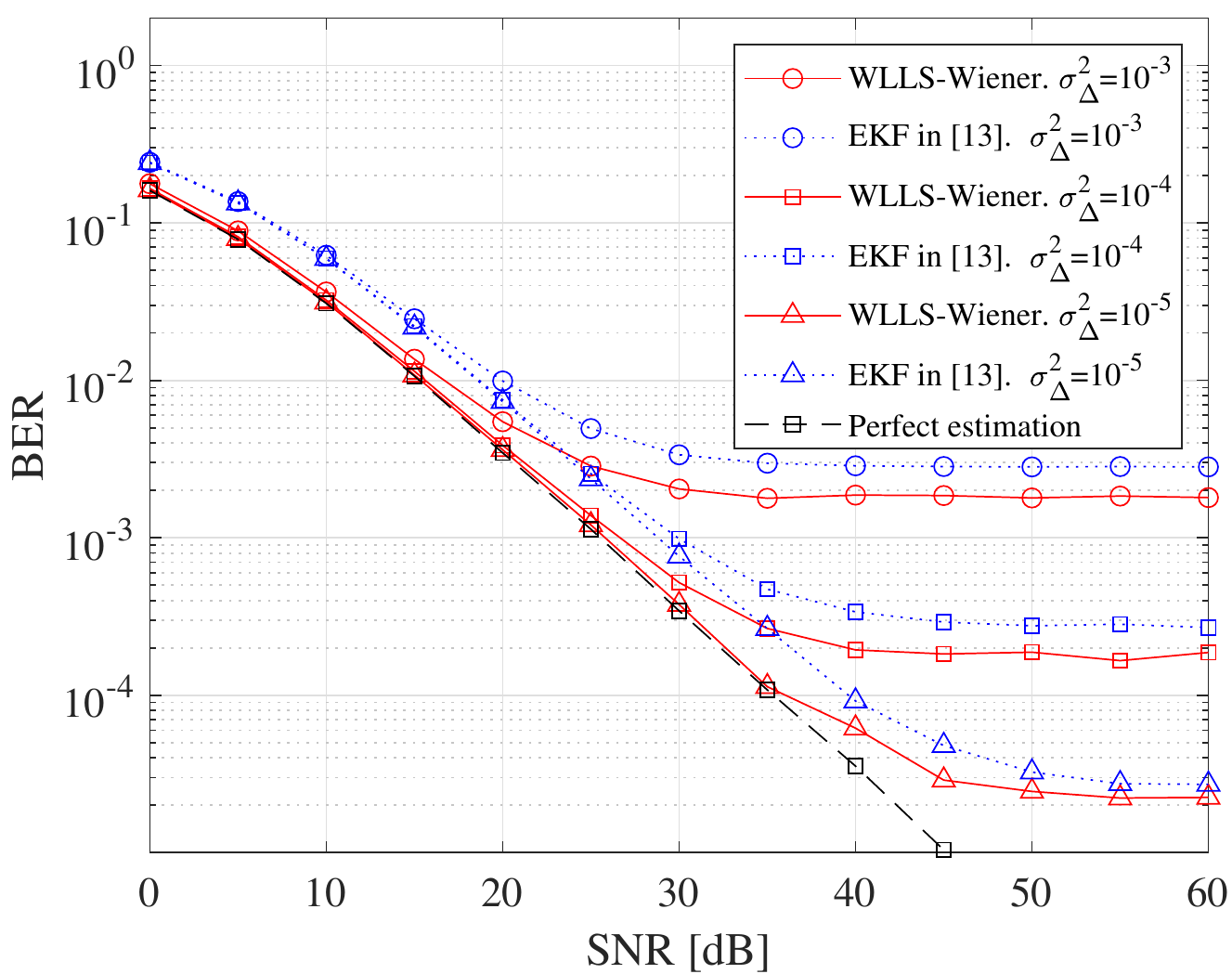}
  \caption{BER of a \(2 \times 2\) MIMO system. \(\sigma _\Delta ^2 = 10^{-3}\) (circles), \(10^{-4}\) (squares), \(10^{-5}\) (triangles). Red lines - proposed algorithm, blue dotted lines - EKF in \cite{mehrpouyan2012joint}, black dashed line - perfect phase and channel estimation.}
  \label{fig:srBer2}
  \end{figure}

Fig.~\ref{fig:srBer2} compares the BER performance of the proposed phase and channel estimation algorithm for different phase noise variances (\(\sigma _\Delta ^2 =10^{-3}, 10^{-4}, 10^{-5}\)). 
In the low SNR region, although the phase noise dependant MSE performance is observed in Fig.~\ref{fig:srPhase4}, we only observe very small BER penalty in Fig.~\ref{fig:srBer2}. This is because the dominant factor in this region is AWGN rather than phase estimation error.
In the high SNR region, the BER floor is lower when the phase noise is smaller, which is a direct consequence of the MSE floor in Fig.~\ref{fig:srPhase4}

\begin{figure}[t]
  \centering
  \includegraphics[width=3.4in]{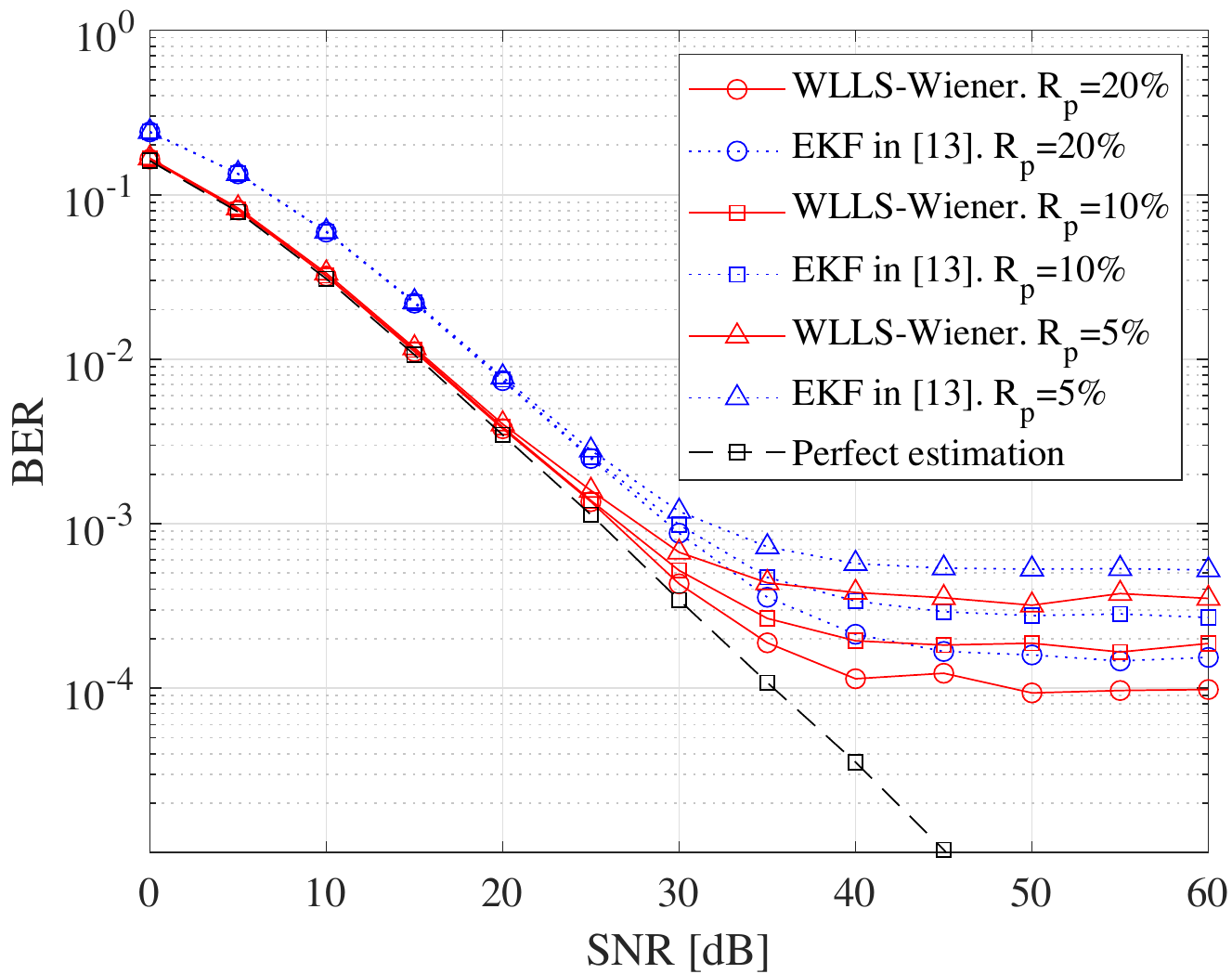}
  \caption{BER of a \(2 \times 2\) MIMO system. Pilot rate \({R_p}=5\%\) (triangles), \(10\%\) (squares), \(20\%\) (circles). Red lines - proposed algorithm, blue dotted lines - EKF in \cite{mehrpouyan2012joint}, black dashed line - perfect phase and channel estimation.}
  \label{fig:srBer3}
  \end{figure}

Fig.~\ref{fig:srBer3} compares the BER performance of the proposed phase and channel estimation algorithm for different pilot rates (\({R_p} =5\%, 10\%, 20\%\)). In the high SNR region, the BER floor slightly decreases when the pilot rate increases. We believe this is mainly because of
the linear interpolation algorithm given by \eqref{equ:36}, which has a better estimation for the data symbols when there are more frequent pilot groups.

\begin{figure}[t]
  \centering
  \includegraphics[width=3.4in]{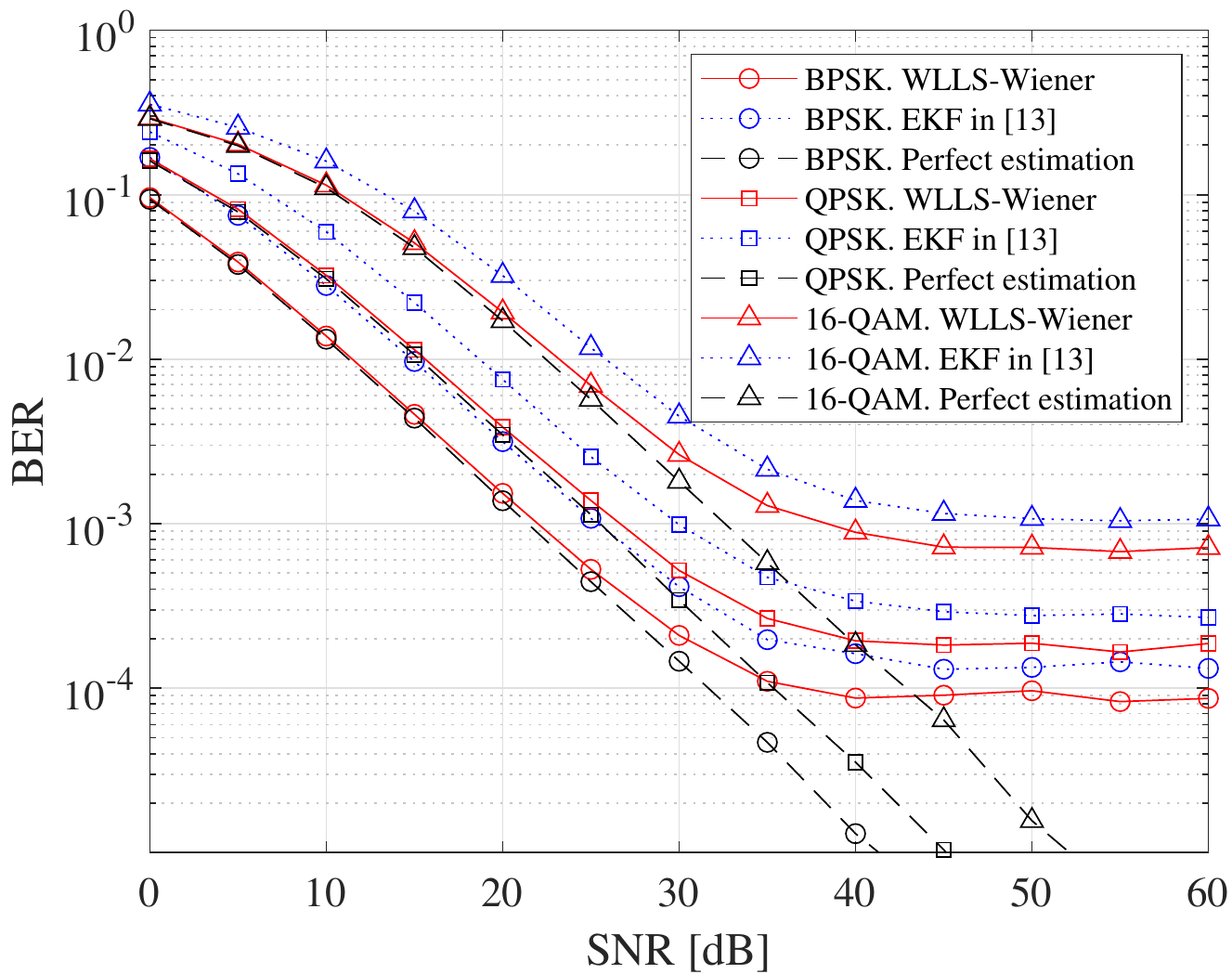}
  \caption{BER of a \(2 \times 2\) MIMO system. Different modulation formats (BPSK (circles), QPSK (squares), and 16-QAM (triangles)). Red lines - proposed algorithm, blue dotted lines - EKF in \cite{mehrpouyan2012joint}, black dashed lines - perfect phase and channel estimation.}
  \label{fig:srBer4}
  \end{figure}

Fig.~\ref{fig:srBer4} compares the BER performance of the proposed phase and channel estimation algorithm for different modulation formats (BPSK, QPSK, and 16-quadrature amplitude modulation (16-QAM)). A higher modulation format is more vulnerable to the phase error, leading to a higher BER floor. As a result, when compared to the perfect phase and channel estimation scenario at the HD-FEC limit of \(\rm{BER} = 4.7 \times 10^{-3}\) \cite{alvarado2015replacing}, the proposed algorithm has an SNR penalty of approximately 0.3~dB, 0.5~dB, and 1.2~dB, while the conventional algorithm has an SNR penalty of approximately 3.5~dB, 3.5~dB, and 4.0~dB for BPSK, QPSK, and 16-QAM, respectively.

\begin{figure}[t]
  \centering
  \includegraphics[width=3.4in]{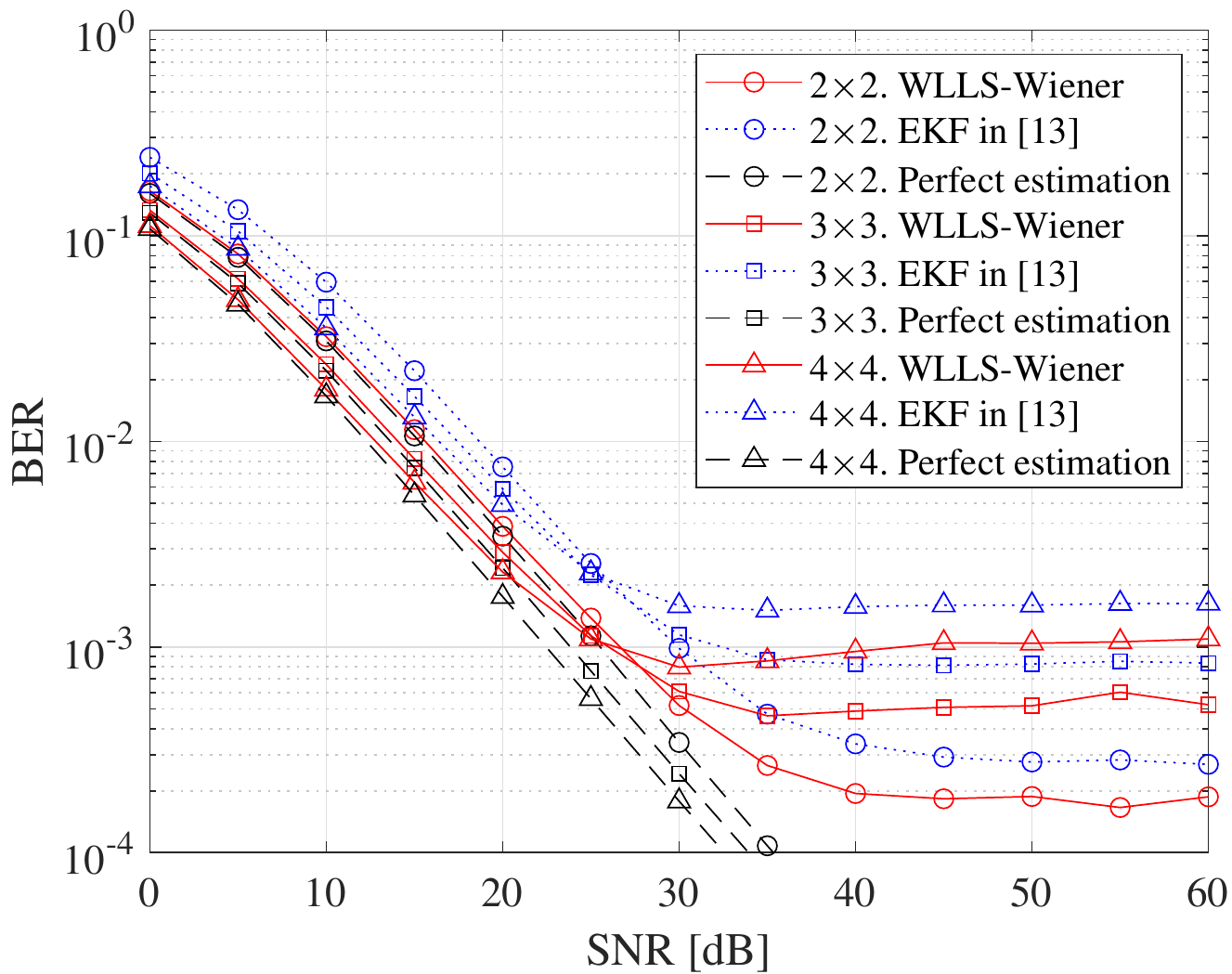}
  \caption{BER of different MIMO systems. \(2\times 2\) (circles), \(3\times 3\) (squares), and \(4\times 4\) (triangles)). Red lines - proposed algorithm, blue dotted lines - EKF in \cite{mehrpouyan2012joint}, black dashed lines - perfect phase and channel estimation.}
  \label{fig:srBer5}
  \end{figure}

Fig.~\ref{fig:srBer5} compares the BER performance of the proposed phase and channel estimation algorithm for different MIMO systems (\(2\times 2\), \(3\times 3\), and \(4 \times 4\)). The results indicates that the larger MIMO scale can result in a higher BER floor. 
This is mainly due to assumption (A1). In order to guarantee orthogonal pilots, the pilot length (\(L_p\)) increases when the transmit antenna number (\(N_t\)) increases. Therefore, when the pilot rate is fixed, a larger transmit antenna number (\(N_t\)) leads to a larger cell length (\({L_c}\)), which results in a worse phase estimation accuracy for the data symbols.
  
\begin{figure}[t]
  \centering
  \includegraphics[width=3.4in]{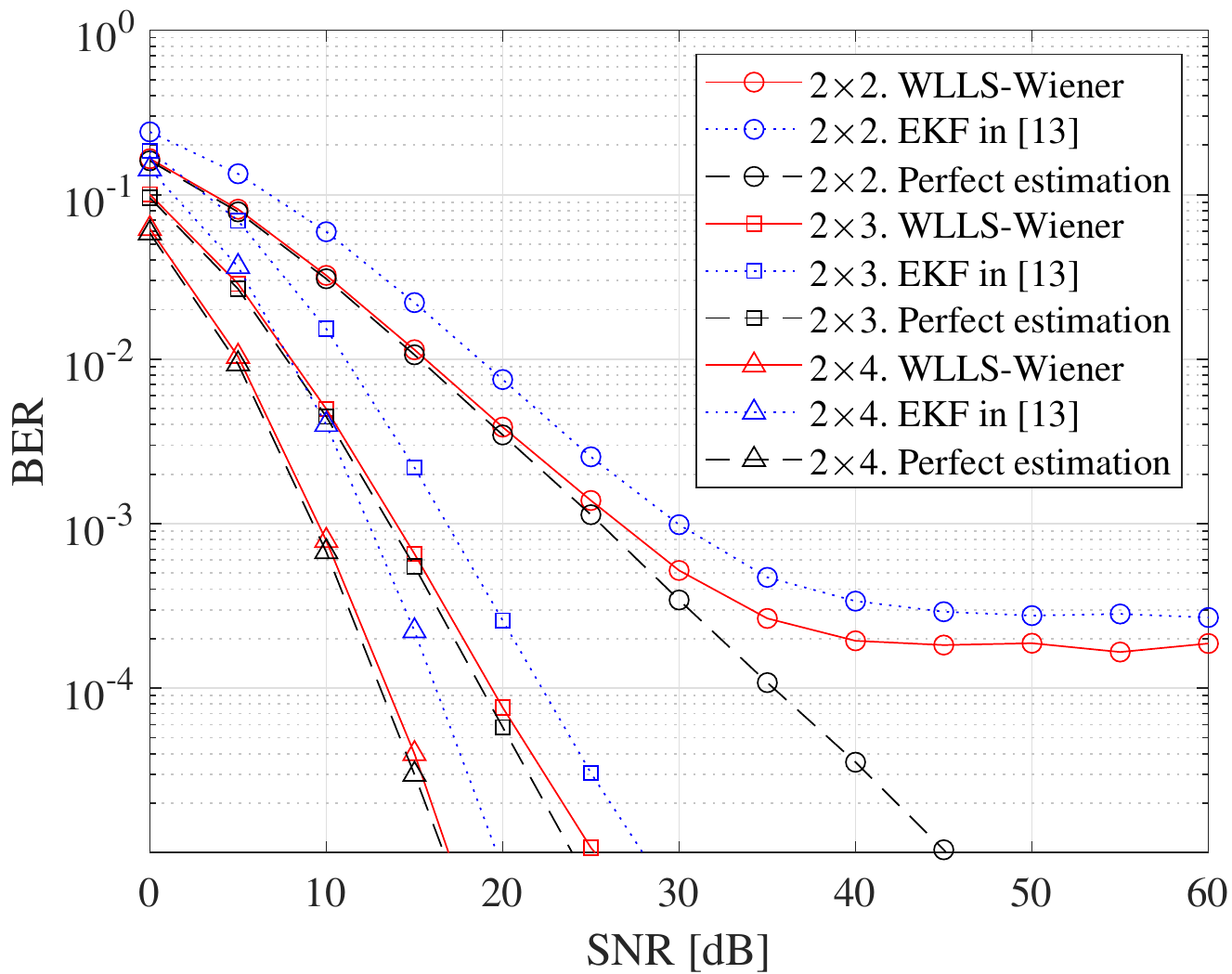}
  \caption{BER of different MIMO systems. \(2\times 2\) (circles), \(2\times 3\) (squares), and \(2\times 4\) (triangles). Red lines - proposed algorithm, blue dotted lines - EKF in \cite{mehrpouyan2012joint}, black dashed lines - perfect phase and channel estimation.}
  \label{fig:srBer6}
  \end{figure}

Fig.~\ref{fig:srBer6} compares the BER performance of the proposed phase and channel estimation algorithm for different MIMO systems (\(2\times 2\), \(2\times 3\), and \(2 \times 4\). The results indicates that for a fixed transmit antenna number (\(N_t\)), a larger receive antenna number (\(N_r\)) can lead to a better BER performance.
This is because the diversity order of MMSE MIMO decoder is \(N_r - N_t + 1\), and a larger diversity order leads to a better performance for the reference system with perfect estimation \cite{Proakis2008communications}.
Moreover, a lower BER floor is also observed when \(N_r\) is larger. And we believe redundant degrees of freedom in MIMO systems can lead to better resistance to the imperfect phase and channel estimations.
As a result, when compared to the perfect phase and channel estimation scenario at the HD-FEC limit of \(\rm{BER} = 4.7 \times 10^{-3}\) \cite{alvarado2015replacing}, the proposed algorithm has an SNR penalty of approximately 0.5~dB, 0.3~dB, and 0.2~dB, while the conventional algorithm has an SNR penalty of approximately 3.6~dB, 3.2~dB, and 3.3~dB for \(2\times 2\), \(2\times 3\), and \(2 \times 4\) MIMO systems, respectively.

\section{Conclusion}
\label{sec:c}
A novel pilot-aided phase and channel estimator is proposed in this paper. The proposed estimator is a sequential combination of a channel amplitude estimator, a WLLS phase estimator, a Wiener phase estimator and an LS channel estimator. 
By doing so, the overall impact of the interaction between phase noise and channel estimation reduces to a quasi-static phase shift, which cancels out when the overall phase and channel estimation is computed. Moreover, the proposed algorithm enables the averaging process for the channel estimation over the whole frame.
As a result, the phase and channel estimation accuracy has been significantly improved. And the phase estimation MSE is very close to the corresponding CRLB. Moreover, at the HD-FEC limit of \(4.7 \times 10^{-3}\), the SNR penalty of the BER curve is reduced to approximately 0.5~dB level, which is more than 2~dB better than the conventional EKF/EKS approaches in \cite{mehrpouyan2012joint, nasir2013phase}.
Furthermore, the algorithm is a sequential combination of several linear estimators, which significantly reduces the computational complexity of the estimation process.

As shown from the numerical results, the proposed algorithm is suitable for different kinds of MIMO decoders (e.g. MMSE and MLD), different modulation formats (e.g. BPSK, QPSK, and 16-QAM), and different antenna numbers.
Although a BER floor is observed in MMSE MIMO decoders when \(N_t = N_r\), this phenomenon can be easily suppressed by using slightly larger scale of receive antennas (\(N_r \ge N_t + 1\)), better decoders such as MLD, or FEC coding, which are typical techniques in commercial systems.

Moreover, when a single oscillator is used in the transmitter or the receiver, which is a special case of the considered system, we can adapt to such cases by slightly modifying the WLLS estimator in the proposed algorithm. As there are fewer phase variables to be estimated, further performance improvement and reduced computational complexity may be achieved.

Considering the performance, the computational complexity, and the adaptability of the newly proposed algorithm, it will provide useful guidelines for designing MIMO systems with phase noise, and it will be feasible to practical commercial systems in the future.


%

\appendices
\section{The Angular Terms of the Observed Data} \label{Appendix:A}
Consider the observed data below
  \begin{equation}\label{equ:A.1}
    {\bf{O}}' = \frac{1}{{{N_t}}}{\bf{Y}}{{\bf{S}}^H}.
  \end{equation}
The element at the \(k^{th}\) row and \(l^{th}\) column of \(\bf{O}'\) can be calculated by \eqref{equ:crlb.05} as
  \begin{equation}\label{equ:A.2}
    \begin{aligned}
    {{o}'_{k,l}} = & \frac{1}{{{N_t}}}{{\bf{y}}_{k,:}}{\bf{s}}_{l,:}^H\\
     \approx & \frac{1}{{{N_t}}}\sum\limits_{l' = 1}^{{N_t}} {{h_{k,l'}}{e^{j\left( {{\varphi _{k,{m_i}}} + {\psi _{l',{m_i}}}} \right)}}{{\bf{s}}_{l',:}}{\bf{s}}_{l,:}^H}  + \frac{1}{{{N_t}}}{{\bf{n}}_{k,:}}{\bf{s}}_{l,:}^H\\
     & + j\frac{1}{{{N_t}}}\sum\limits_{l' = 1}^{{N_t}} {{h_{k,l'}}{e^{j\left( {{\varphi _{k,{m_i}}} + {\psi _{l',{m_i}}}} \right)}}\left( {{{\bf{s}}_{l',:}} \odot {{\boldsymbol{\gamma }}_{k,l',:}}} \right){\bf{s}}_{l,:}^H} .
    \end{aligned}
  \end{equation}
Note the fact that \(\bf{S}\) is an orthogonal matrix with normalized elements, \eqref{equ:A.2} can be simplified and rewritten in element-wise form as
  \begin{equation}\label{equ:A.3}
    \begin{aligned}
    {{o}'_{k,l}} & \approx {h_{k,l}}{e^{j\left( {{\varphi _{k,{m_i}}} + {\psi _{l,{m_i}}}} \right)}} + n_{k,l}^{\left( 1 \right)}\\
     & + j\sum\limits_{l' = 1}^{{N_t}} {\left[ {\frac{{{h_{k,l'}}{e^{j\left( {{\varphi _{k,{m_i}}} + {\psi _{l',{m_i}}}} \right)}}}}{{{N_t}}}\sum\limits_{m = 1}^{{N_t}} {\left( {{s_{l',m}}s_{l,m}^*{\gamma _{k,l',m}}} \right)} } \right]} ,
    \end{aligned}
  \end{equation}
where \(n_{k,l}^{\left( 1 \right)} \sim {\cal C}{\cal N}\left( {0,{{\sigma _n^2} \mathord{\left/
 {\vphantom {{\sigma _n^2} {{N_t}}}} \right.
 \kern-\nulldelimiterspace} {{N_t}}}} \right)\) is i.i.d. circularly-symmetric complex AWGN.

Define
  \begin{equation}\label{equ:A.4}
   {\eta _{k,l,l',m}} \buildrel \Delta \over = \frac{{{h_{k,l'}}}}{{{N_t}{h_{k,l}}}}{e^{j\left( {{\psi _{l',{m_i}}} - {\psi _{l,{m_i}}}} \right)}}{s_{l',m}}s_{l,m}^* .
  \end{equation}
Equation \eqref{equ:A.3} can be represented as
  \begin{equation}\label{equ:A.5}
    \begin{aligned}
    {{o}'_{k,l}} & \approx {h_{k,l}}{e^{j\left( {{\varphi _{k,{m_i}}} + {\psi _{l,{m_i}}}} \right)}}\\
     & + {h_{k,l}}{e^{j\left( {{\varphi _{k,{m_i}}} + {\psi _{l,{m_i}}}} \right)}} \cdot \frac{{n_{k,l}^{\left( 1 \right)}}}{{{h_{k,l}}{e^{j\left( {{\varphi _{k,{m_i}}} + {\psi _{l,{m_i}}}} \right)}}}}\\
     & + {h_{k,l}}{e^{j\left( {{\varphi _{k,{m_i}}} + {\psi _{l,{m_i}}}} \right)}}j\sum\limits_{l' = 1}^{{N_t}} {\sum\limits_{m = 1}^{{N_t}} {{\eta _{k,l,l',m}}{\gamma _{k,l',m}}} } .
    \end{aligned}
  \end{equation}
  
When the complex term \(X\) is assumed to be small, below approximation holds
  \begin{equation}\label{equ:A.6}
    1 + X \approx \left( {1 + \Re \left( X \right)} \right){e^{j\Im \left( X \right)}} .
  \end{equation}
By extracting the term \({h_{k,l}}{e^{j\left( {{\varphi _{k,{m_i}}} + {\psi _{l,{m_i}}}} \right)}}\) in \eqref{equ:A.5} and using \eqref{equ:A.6}, \eqref{equ:A.5} can be rewritten as
  \begin{equation}\label{equ:A.7}
    \begin{aligned}
    {{o}'_{k,l}} & \approx {h_{k,l}}\left( {1 + \frac{{n_{k,l}^{\left( 3 \right)}}}{{\left| {{h_{k,l}}} \right|}} + \sum\limits_{l' = 1}^{{N_t}} {\sum\limits_{m = 1}^{{N_t}} {\Im \left( {{\eta _{k,l,l',m}}} \right){\gamma _{k,l',m}}} } } \right)\\
     & \times {e^{j\left( {{\varphi _{k,{m_i}}} + {\psi _{l,{m_i}}} + \frac{{n_{k,l}^{\left( 2 \right)}}}{{\left| {{h_{k,l}}} \right|}} + \sum\limits_{l' = 1}^{{N_t}} {\sum\limits_{m = 1}^{{N_t}} {\Re \left( {{\eta _{k,l,l',m}}} \right){\gamma _{k,l',m}}} } } \right)}} ,
    \end{aligned}
  \end{equation}
where
  \begin{equation}\label{equ:A.8}
    \left\{ \begin{aligned}
    n_{k,l}^{\left( 2 \right)} = & \Im \left( {n_{k,l}^{\left( 1 \right)}{e^{ - j\left( {\angle {h_{k,l}} + {\varphi _{k,{m_i}}} + {\psi _{l,{m_i}}}} \right)}}} \right)\\
    n_{k,l}^{\left( 3 \right)} = & \Re \left( {n_{k,l}^{\left( 1 \right)}{e^{ - j\left( {\angle {h_{k,l}} + {\varphi _{k,{m_i}}} + {\psi _{l,{m_i}}}} \right)}}} \right) .
    \end{aligned} \right.
  \end{equation}
are i.i.d. real AWGN with zero mean and variance \({{{\sigma _n^2} \mathord{\left/
 {\vphantom {{\sigma _n^2} {2{N_t}}}} \right.
 \kern-\nulldelimiterspace} {2{N_t}}}}\).

Noting the fact that \(n_{k,l}^{\left( 1 \right)}\) is i.i.d. circularly-symmetric complex AWGN, the exponential term in \eqref{equ:A.8} does not change the distribution of \(n_{k,l}^{\left( 2 \right)}\) and \(n_{k,l}^{\left( 3 \right)}\). Therefore, the information of \({\varphi _{k,m_i}}\) and \({\psi _{l,m_i}}\) is only included in the angular term of \eqref{equ:A.7}. If perfect channel estimation is assumed, which is the optimal case of the phase estimation, the effective observed data for the phase estimation is the angular term of \eqref{equ:A.7}, which can be written as
  \begin{equation}\label{equ:A.9}
    {o_{k,l}} = {\varphi _{k,{m_i}}} + {\psi _{l,{m_i}}} + \frac{{n_{k,l}^{\left( 2 \right)}}}{{\left| {{h_{k,l}}} \right|}} + \sum\limits_{l' = 1}^{{N_t}} {\sum\limits_{m = 1}^{{N_t}} {{\xi _{k,l,l',m}}{\gamma _{k,l',m}}} } ,
  \end{equation}
where \({\xi _{k,l,l',m}} = {\Re \left( {{\eta _{k,l,l',m}}} \right)}\) is given by \eqref{equ:crlb.07}.

By substituting \eqref{equ:crlb.02} into \eqref{equ:A.9}, the equation can be rewritten as
  \begin{equation}\label{equ:A.10}
    \begin{aligned}
    {o_{k,l}} = & {\varphi _{k,{m_i}}} + {\psi _{l,{m_i}}} + \frac{{n_{k,l}^{\left( 2 \right)}}}{{\left| {{h_{k,l}}} \right|}}\\
     & - \sum\limits_{l' = 1}^{{N_t}} {\sum\limits_{m = 1}^{{m_i} - 1} {\sum\limits_{m' = m + 1}^{{m_i}} {{\xi _{k,l,l',m}}\left( {\Delta {\varphi _{k,m'}} + \Delta {\psi _{l',m'}}} \right)} } } \\
     & + \sum\limits_{l' = 1}^{{N_t}} {\sum\limits_{m = {m_i} + 1}^{{N_t}} {\sum\limits_{m' = {m_i} + 1}^m {{\xi _{k,l,l',m}}\left( {\Delta {\varphi _{k,m'}} + \Delta {\psi _{l',m'}}} \right)} } } \\
     = & {\varphi _{k,{m_i}}} + {\psi _{l,{m_i}}} + \frac{{n_{k,l}^{\left( 2 \right)}}}{{\left| {{h_{k,l}}} \right|}}\\
     & - \sum\limits_{l' = 1}^{{N_t}} {\sum\limits_{m = 1}^{{m_i} - 1} {\sum\limits_{m' = m + 1}^{{m_i}} {{\xi _{k,l,l',m}}\Delta {\varphi _{k,m'}}} } } \\
     & - \sum\limits_{l' = 1}^{{N_t}} {\sum\limits_{m = 1}^{{m_i} - 1} {\sum\limits_{m' = m + 1}^{{m_i}} {{\xi _{k,l,l',m}}\Delta {\psi _{l',m'}}} } } \\
     & + \sum\limits_{l' = 1}^{{N_t}} {\sum\limits_{m = {m_i} + 1}^{{N_t}} {\sum\limits_{m' = {m_i} + 1}^m {{\xi _{k,l,l',m}}\Delta {\varphi _{k,m'}}} } } \\
     & + \sum\limits_{l' = 1}^{{N_t}} {\sum\limits_{m = {m_i} + 1}^{{N_t}} {\sum\limits_{m' = {m_i} + 1}^m {{\xi _{k,l,l',m}}\Delta {\psi _{l',m'}}} } } .
    \end{aligned}
  \end{equation}
And \eqref{equ:crlb.06} can be directly obtained by changing the order of summation in \eqref{equ:A.10}. 

\section{Derivation of \eqref{equ:crlb.13}}
\label{Appendix:B}

Equation \eqref{equ:crlb.12} can be rewritten in element-wise form as
  \begin{equation}\label{equ:B.1}
    \begin{gathered}
      {\Sigma _{\upsilon \left( {{N_t}\left( {{k_1} - 1} \right) + {l_1},{N_t}\left( {{k_2} - 1} \right) + {l_2}} \right)}} \hfill \\
       = {\text{E}}\left[ {\left( {{o_{{k_1},{l_1}}} - {\mu _{\upsilon \left( {{N_t}\left( {{k_1} - 1} \right) + {l_1}} \right)}}} \right)\left( {{o_{{k_2},{l_2}}} - {\mu _{\upsilon \left( {{N_t}\left( {{k_2} - 1} \right) + {l_2}} \right)}}} \right)} \right] . \hfill \\ 
    \end{gathered}
  \end{equation}

Noting the fact that all the phase noise increment and AWGN are mutually independent random variables, the cross-covariance between any two different noise sources is zero. By substituting \eqref{equ:crlb.06} and \eqref{equ:crlb.10} into \eqref{equ:B.1}, the \(1^{st}\) non-zero term of \eqref{equ:B.1} can be written as
  \begin{equation}\label{equ:B.2}
    \begin{gathered}
      {\text{E}}\left( {\frac{{n_{{k_1},{l_1}}^{\left( 2 \right)}}}{{\left| {{h_{{k_1},{l_1}}}} \right|}}\frac{{n_{{k_2},{l_2}}^{\left( 2 \right)}}}{{\left| {{h_{{k_2},{l_2}}}} \right|}}} \right) \hfill \\
       = \frac{1}{{\left| {{h_{{k_1},{l_1}}}} \right|\left| {{h_{{k_2},{l_2}}}} \right|}}{\text{E}}\left( {n_{{k_1},{l_1}}^{\left( 2 \right)}n_{{k_2},{l_2}}^{\left( 2 \right)}} \right) \hfill \\
       = \frac{1}{{\left| {{h_{{k_1},{l_1}}}} \right|\left| {{h_{{k_2},{l_2}}}} \right|}} \cdot \frac{{\sigma _n^2}}{{2{N_t}}}\delta \left( {{k_1} - {k_2}} \right)\delta \left( {{k_l} - {k_l}} \right) \hfill \\
       = \frac{1}{{2{N_t}\left| {{h_{{k_1},{l_1}}}} \right|\left| {{h_{{k_2},{l_2}}}} \right|}}\sigma _n^2\delta \left( {{k_1} - {k_2}} \right)\delta \left( {{k_l} - {k_l}} \right) . \hfill \\ 
    \end{gathered}
  \end{equation}

The \(2^{nd}\) non-zero term of \eqref{equ:B.1} can be written as
  \begin{equation}\label{equ:B.3}
    \begin{gathered}
      {\text{E}} \left( { - \sum\limits_{{m_1'} = 2}^{{m_i}} {\left( {\sum\limits_{{m_1} = 1}^{{m_1'} - 1} {\sum\limits_{{l_1'} = 1}^{{N_t}} {{\xi _{{k_1},{l_1},{l_1'},{m_1}}}} } } \right)\Delta {\varphi _{{k_1},{m_1'}}}} } \right. \hfill \\
      \left. { \times \left[ { - \sum\limits_{{m_2'} = 2}^{{m_i}} {\left( {\sum\limits_{{m_2} = 1}^{{m_2'} - 1} {\sum\limits_{{l_2'} = 1}^{{N_t}} {{\xi _{{k_2},{l_2},{l_2'},{m_2}}}} } } \right)\Delta {\varphi _{{k_2},{m_2'}}}} } \right]} \right) \hfill \\
       = \sum\limits_{m' = 2}^{{m_i}} {\left[ {\left( {\sum\limits_{{m_1} = 1}^{m' - 1} {\sum\limits_{{l_1'} = 1}^{{N_t}} {{\xi _{{k_1},{l_1},{l_1'},{m_1}}}} } } \right)} \right.}  \hfill \\
      \left. { \times \left( {\sum\limits_{{m_2} = 1}^{m' - 1} {\sum\limits_{{l_2'} = 1}^{{N_t}} {{\xi _{{k_2},{l_2},{l_2'},{m_2}}}} } } \right){\text{E}}\left( {\Delta {\varphi _{{k_1},m'}}\Delta {\varphi _{{k_2},m'}}} \right)} \right] \hfill \\
       = \sum\limits_{m' = 2}^{{m_i}} {\left[ {\left( {\sum\limits_{{m_1} = 1}^{m' - 1} {\sum\limits_{{l_1'} = 1}^{{N_t}} {{\xi _{{k_1},{l_1},{l_1'},{m_1}}}} } } \right)} \right.}  \hfill \\
      \left. { \times \left( {\sum\limits_{{m_2} = 1}^{m' - 1} {\sum\limits_{{l_2'} = 1}^{{N_t}} {{\xi _{{k_2},{l_2},{l_2'},{m_2}}}} } } \right)\sigma _{\Delta \varphi }^2\delta \left( {{k_1} - {k_2}} \right)} \right]. \hfill \\ 
    \end{gathered}
  \end{equation}
The simplification of the first and second equation in \eqref{equ:B.3} is due to the fact that
  \begin{equation}\label{equ:B.4}
    \begin{aligned}
    & {\text{E}} \left( {\Delta {\varphi _{{k_1},{m_1'}}}\Delta {\varphi _{{k_2},{m_2'}}}} \right)\\
    &  = \left\{ {\begin{aligned}
    & \sigma _{\Delta \varphi }^2, & \left( {{k_1} = {k_2},{m_1'} = {m_2'} = m'} \right) &\\
    & 0, & \left( {otherwise} \right)  & .
    \end{aligned}} \right.
    \end{aligned}
  \end{equation}
  
The \(3^{rd}\) non-zero term of \eqref{equ:B.1} can be written as
  \begin{equation}\label{equ:B.5}
    \begin{aligned}
    & {\rm{E}}\left( {\left[ { - \sum\limits_{{l_1'} = 1}^{{N_t}} {\sum\limits_{{m_1'} = 2}^{{m_i}} {\left( {\sum\limits_{{m_1} = 1}^{{m_1'} - 1} {{\xi _{{k_1},{l_1},{l_1'},{m_1}}}} } \right)\Delta {\psi _{{l_1'},{m_1'}}}} } } \right]} \right.\\
    & \left. { \times \left[ { - \sum\limits_{{l_2'} = 1}^{{N_t}} {\sum\limits_{{m_2'} = 2}^{{m_i}} {\left( {\sum\limits_{{m_2} = 1}^{{m_2'} - 1} {{\xi _{{k_2},{l_2},{l_2'},{m_2}}}} } \right)\Delta {\psi _{{l_2'},{m_2'}}}} } } \right]} \right)\\
    & = \sum\limits_{l' = 1}^{{N_t}} {\sum\limits_{m' = 2}^{{m_i}} {\left[ {\left( {\sum\limits_{{m_1} = 1}^{{m_1'} - 1} {{\xi _{{k_1},{l_1},{l_1'},{m_1}}}} } \right)} \right.} } \\
    & \left. { \times \left( {\sum\limits_{{m_2} = 1}^{{m_2'} - 1} {{\xi _{{k_2},{l_2},{l_2'},{m_2}}}} } \right)\sigma _{\Delta \psi }^2} \right].
    \end{aligned}
  \end{equation}
      
The simplification of \eqref{equ:B.5} is due to the fact that 
  \begin{equation}\label{equ:B.6}
    \begin{aligned}
    & {\text{E}} \left( {\Delta {\psi _{{l_1'},{m_1'}}}\Delta {\psi _{{l_2'},{m_2'}}}} \right)\\
    & = \left\{ {\begin{aligned}
    & \sigma _{\Delta \psi }^2, & \left( {{l_1'} = {l_2'} = l',{m_1'} = {m_2'} = m'} \right) & \\
    & 0, & \left( {otherwise} \right) &.
    \end{aligned}} \right.
    \end{aligned}
  \end{equation}

Similar to \eqref{equ:B.3}, the \(4^{th}\) non-zero term of \eqref{equ:B.1} can be written as
  \begin{equation}\label{equ:B.7}
    \begin{aligned}
    & {\rm{E}}\left[ {\sum\limits_{{m_1'} = {m_i} + 1}^{{N_t}} {\left( {\sum\limits_{{m_1} = {m_1'}}^{{N_t}} {\sum\limits_{{l_1'} = 1}^{{N_t}} {{\xi _{{k_1},{l_1},{l_1'},{m_1}}}} } } \right)\Delta {\varphi _{{k_1},{m_1'}}}} } \right.\\
    & \left. { \times \sum\limits_{{m_2'} = {m_i} + 1}^{{N_t}} {\left( {\sum\limits_{{m_2} = {m_2'}}^{{N_t}} {\sum\limits_{{l_2'} = 1}^{{N_t}} {{\xi _{{k_2},{l_2},{l_2'},{m_2}}}} } } \right)\Delta {\varphi _{{k_2},{m_2'}}}} } \right]\\
    & = \sum\limits_{m' = {m_i} + 1}^{{N_t}} {\left[ {\left( {\sum\limits_{{m_1} = m'}^{{N_t}} {\sum\limits_{{l_1'} = 1}^{{N_t}} {{\xi _{{k_1},{l_1},{l_1'},{m_1}}}} } } \right)} \right.} \\
    & \left. { \times \left( {\sum\limits_{{m_2} = m'}^{{N_t}} {\sum\limits_{{l_2'} = 1}^{{N_t}} {{\xi _{{k_2},{l_2},{l_2'},{m_2}}}} } } \right)\sigma _{\Delta \varphi }^2\delta \left( {{k_1} - {k_2}} \right)} \right].
    \end{aligned}
  \end{equation}

Similar to \eqref{equ:B.5}, the \(5^{th}\) non-zero term of \eqref{equ:B.1} can be written as
  \begin{equation}\label{equ:B.8}
    \begin{aligned}
    & {\rm{E}}\left( {\left[ {\sum\limits_{{l_1'} = 1}^{{N_t}} {\sum\limits_{{m_1'} = {m_i} + 1}^{{N_t}} {\left( {\sum\limits_{{m_1} = {m_1'}}^{{N_t}} {{\xi _{{k_1},{l_1},{l_1'},{m_1}}}} } \right)\Delta {\psi _{{l_1'},{m_1'}}}} } } \right]} \right.\\
    & \left. { \times \left[ {\sum\limits_{{l_2'} = 1}^{{N_t}} {\sum\limits_{{m_2'} = {m_i} + 1}^{{N_t}} {\left( {\sum\limits_{{m_2} = {m_2'}}^{{N_t}} {{\xi _{{k_2},{l_2},{l_2'},{m_2}}}} } \right)\Delta {\psi _{{l_2'},{m_2'}}}} } } \right]} \right)\\
     & = \sum\limits_{l' = 1}^{{N_t}} {\sum\limits_{m' = {m_i} + 1}^{{N_t}} {\left[ {\left( {\sum\limits_{{m_1} = m'}^{{N_t}} {{\xi _{{k_1},{l_1},l',{m_1}}}} } \right)} \right.} } \\
    & \left. { \times \left( {\sum\limits_{{m_2} = {m_2'}}^{{N_t}} {{\xi _{{k_2},{l_2},{l_2'},{m_2}}}} } \right)\sigma _{\Delta \psi }^2} \right] .
    \end{aligned}
  \end{equation}

Moreover, all the other terms in \eqref{equ:B.1} are equal to \(0\). And \eqref{equ:crlb.13} can be obtained by substituting \eqref{equ:B.2}, \eqref{equ:B.3}, \eqref{equ:B.5}, \eqref{equ:B.7}, \eqref{equ:B.8} into \eqref{equ:B.1}.

\section{Derivaton of \eqref{equ:crlb.15}}
\label{Appendix:C}

By substituting \eqref{equ:19} into \eqref{equ:crlb.07}, the first derivative under different cases can be calculated as below:

When \(l' \ne l\), \(q = l' + {N_r}\), and \(l \ne {N_t}\), the first derivative of \eqref{equ:crlb.07} has the form of
  \begin{equation}\label{equ:C.1}
    \begin{aligned}
    \frac{{\partial {\xi _{k,l,l',m}}}}{{\partial {\beta _{q,{m_i}}}}} = & \frac{\partial }{{\partial {\beta _{q,{m_i}}}}}\Re \left( {\frac{{{h_{k,l'}}}}{{{N_t}{h_{k,l}}}}{e^{j\left( {{\beta _{q,{m_i}}} - {\beta _{l + {N_r},{m_i}}}} \right)}}{s_{l',m}}s_{l,m}^*} \right)\\
     = & \Re \left( {j\frac{{{h_{k,l'}}}}{{{N_t}{h_{k,l}}}}{e^{j\left( {{\beta _{q,{m_i}}} - {\beta _{l + {N_r},{m_i}}}} \right)}}{s_{l',m}}s_{l,m}^*} \right)\\
     = &  - \Im \left( {\frac{{{h_{k,l'}}}}{{{N_t}{h_{k,l}}}}{e^{j\left( {{\beta _{q,{m_i}}} - {\beta _{l + {N_r},{m_i}}}} \right)}}{s_{l',m}}s_{l,m}^*} \right)\\
     = &  - \Im \left( {\frac{{{h_{k,l'}}}}{{{N_t}{h_{k,l}}}}{e^{j\left( {{\psi _{l',{m_i}}} - {\psi _{l,{m_i}}}} \right)}}{s_{l',m}}s_{l,m}^*} \right).
    \end{aligned}
  \end{equation}
  
When \(l' \ne l\), \(q = l' + {N_r}\), and \(l = {N_t}\), the first derivative of \eqref{equ:crlb.07} has the form of
  \begin{equation}\label{equ:C.2}
    \begin{aligned}
    \frac{{\partial {\xi _{k,l,l',m}}}}{{\partial {\beta _{q,{m_i}}}}} & = \frac{\partial }{{\partial {\beta _{q,{m_i}}}}}\Re \left( {\frac{{{h_{k,l'}}}}{{{N_t}{h_{k,l}}}}{e^{j{\beta _{q,{m_i}}}}}{s_{l',m}}s_{l,m}^*} \right)\\
     & = \Re \left( {j\frac{{{h_{k,l'}}}}{{{N_t}{h_{k,l}}}}{e^{j{\beta _{q,{m_i}}}}}{s_{l',m}}s_{l,m}^*} \right)\\
     & =  - \Im \left( {\frac{{{h_{k,l'}}}}{{{N_t}{h_{k,l}}}}{e^{j{\beta _{q,{m_i}}}}}{s_{l',m}}s_{l,m}^*} \right)\\
     & =  - \Im \left( {\frac{{{h_{k,l'}}}}{{{N_t}{h_{k,l}}}}{e^{j\left( {{\psi _{l',{m_i}}} - {\psi _{l,{m_i}}}} \right)}}{s_{l',m}}s_{l,m}^*} \right) .
    \end{aligned}
  \end{equation}

When \(l' \ne l\), \(q = l + {N_r}\), and \(l' \ne {N_t}\), the first derivative of \eqref{equ:crlb.07} has the form of
  \begin{equation}\label{equ:C.3}
    \begin{aligned}
    & \frac{{\partial {\xi _{k,l,l',m}}}}{{\partial {\beta _{q,{m_i}}}}} \\
    & = \frac{\partial }{{\partial {\beta _{q,{m_i}}}}}\Re \left( {\frac{{{h_{k,l'}}}}{{{N_t}{h_{k,l}}}}{e^{j\left( {{\beta _{l' + {N_r},{m_i}}} - {\beta _{q,{m_i}}}} \right)}}{s_{l',m}}s_{l,m}^*} \right)\\
     & = \Re \left( { - j\frac{{{h_{k,l'}}}}{{{N_t}{h_{k,l}}}}{e^{j\left( {{\beta _{l' + {N_r},{m_i}}} - {\beta _{q,{m_i}}}} \right)}}{s_{l',m}}s_{l,m}^*} \right)\\
     & = \Im \left( {\frac{{{h_{k,l'}}}}{{{N_t}{h_{k,l}}}}{e^{j\left( {{\beta _{l' + {N_r},{m_i}}} - {\beta _{q,{m_i}}}} \right)}}{s_{l',m}}s_{l,m}^*} \right)\\
     & = \Im \left( {\frac{{{h_{k,l'}}}}{{{N_t}{h_{k,l}}}}{e^{j\left( {{\psi _{l',{m_i}}} - {\psi _{l,{m_i}}}} \right)}}{s_{l',m}}s_{l,m}^*} \right) .
    \end{aligned}
  \end{equation}

When \(l' \ne l\), \(q = l + {N_r}\), and \(l' = {N_t}\), the first derivative of \eqref{equ:crlb.07} has the form of
  \begin{equation}\label{equ:C.4}
    \begin{aligned}
    \frac{{\partial {\xi _{k,l,l',m}}}}{{\partial {\beta _{q,{m_i}}}}} & = \frac{\partial }{{\partial {\beta _{q,{m_i}}}}}\Re \left( {\frac{{{h_{k,l'}}}}{{{N_t}{h_{k,l}}}}{e^{ - j{\beta _{q,{m_i}}}}}{s_{l',m}}s_{l,m}^*} \right)\\
     & = \Re \left( { - j\frac{{{h_{k,l'}}}}{{{N_t}{h_{k,l}}}}{e^{ - j{\beta _{q,{m_i}}}}}{s_{l',m}}s_{l,m}^*} \right)\\
     & = \Im \left( {\frac{{{h_{k,l'}}}}{{{N_t}{h_{k,l}}}}{e^{ - j{\beta _{q,{m_i}}}}}{s_{l',m}}s_{l,m}^*} \right)\\
     & = \Im \left( {\frac{{{h_{k,l'}}}}{{{N_t}{h_{k,l}}}}{e^{j\left( {{\psi _{l',{m_i}}} - {\psi _{l,{m_i}}}} \right)}}{s_{l',m}}s_{l,m}^*} \right) .
    \end{aligned}
  \end{equation}
  
When \(l' \ne l\), \(q \ne l + {N_r}\), and \(q \ne l' + {N_r}\), the first derivative of \eqref{equ:crlb.07} has the form of
  \begin{equation}\label{equ:C.5}
    \begin{aligned}
    \frac{{\partial {\xi _{k,l,l',m}}}}{{\partial {\beta _{q,{m_i}}}}} & = \frac{\partial }{{\partial {\beta _{q,{m_i}}}}}\Re \left( {\frac{{{h_{k,l'}}}}{{{N_t}{h_{k,l}}}}{e^{j\left( {{\psi _{l',{m_i}}} - {\psi _{l,{m_i}}}} \right)}}{s_{l',m}}s_{l,m}^*} \right)\\
    & = 0 .
    \end{aligned}
  \end{equation}

When \(l' = l\), the first derivative of \eqref{equ:crlb.07} has the form of
  \begin{equation}\label{equ:C.6}
    \frac{{\partial {\xi _{k,l,l',m}}}}{{\partial {\beta _{q,{m_i}}}}} = \frac{\partial }{{\partial {\beta _{q,{m_i}}}}}\Re \left( {\frac{{{h_{k,l'}}}}{{{N_t}{h_{k,l}}}}{s_{l',m}}s_{l,m}^*} \right) = 0 .
  \end{equation}

And \eqref{equ:crlb.15} is obtained by directly combining \eqref{equ:C.1}-\eqref{equ:C.6}.

\section*{Acknowledgment}

The authors would like to thank Dr. Chao Gao and Dr. Long Jian for fruitful discussions on the mathematical derivation.











\ifCLASSOPTIONcaptionsoff
  \newpage
\fi



\bibliographystyle{IEEEtran}
\bibliography{paperbib}

%




\begin{IEEEbiography}[{\includegraphics[width=1in,height=1.25in,clip,keepaspectratio]{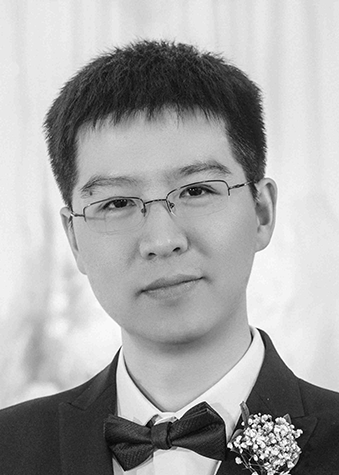}}]{Yiming Li} received the B.S. degree, the M.S. degree and the Ph.D. degree, all in electrical engineering, from the University of Electronic Science and Technology of China, China, in 2011, 2014 and 2019, respectively.

He is now a Postdoctoral Researcher (Marie-S Curie Research Fellow) at Aston Insitute of Photonic Technology (AiPT), Aston University. His current research interests lie in the area of wireless communication systems and signal processing, including MIMO decoding, channel estimation, and performance optimization.

Dr. Li has received more than 10 scholarships and awards, e.g., Marie-S Curie Grant, Chinese National Scholarship, Lixin Tang Scholarship, etc.
\end{IEEEbiography}

\begin{IEEEbiography}[{\includegraphics[width=1in,height=1.25in,clip,keepaspectratio]{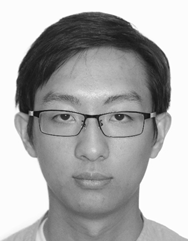}}]{Zhouyi Hu} (Member, IEEE) received the B.S. degree in optoelectronic information engineering from Huazhong University of Science and Technology in 2016 and the Ph.D. degree in information engineering from The Chinese University of Hong Kong in 2020. From July 2019 to April 2020, he was also with University College London as a visiting researcher. Upon graduation, he worked as a research associate at The Chinese University of Hong Kong from October 2020 to November 2020. Since December 2020, he is a research associate at AiPT, Aston University.

His research interests include physical layer security in PONs, optical wireless communications, advanced modulation formats, and DSP for optical systems.
\end{IEEEbiography}

\begin{IEEEbiography}[{\includegraphics[width=1in,height=1.25in,clip,keepaspectratio]{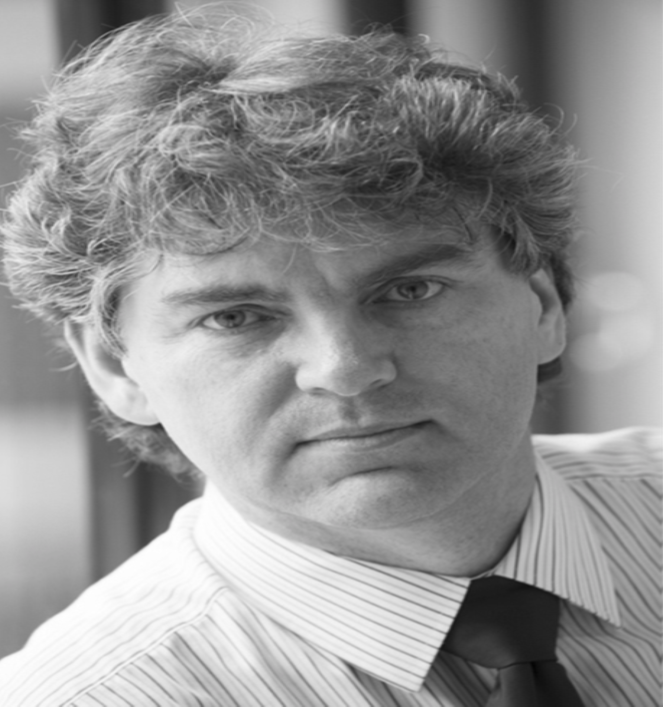}}]{Andrew D. Ellis} was born in Underwood, U.K., in 1965. He received the B.Sc. degree in physics with a minor in mathematics from the University of Sussex, Brighton, U.K., in 1987. He received the Ph.D. degree in electronic and electrical engineering from The University of Aston in Birmingham, Birmingham, U.K., in 1997 for his study on all optical networking beyond 10 Gbit/s.

He previously worked for British Telecom Research Laboratories as a Senior Research Engineer investigating the use of optical amplifiers and advanced modulation formats in optical networks and the Corning Research Centre as a Senior Research Fellow where he led activities in optical component characterization. From 2003, he headed the Transmission and Sensors Group at the Tyndall National Institute in Cork, Ireland, where he was also a member of the Department of Physics, University College Cork and his research interests included the evolution of core and metro networks, and the application of photonics to sensing. He is now 50th Anniversary Professor of Optical Communications at Aston University where he is also deputy director of the Institute of Photonics Technologies (AiPT) where he is continuing his research to increase the reach, capacity and functionality of optical networks. He has published over 200 journal papers and over 28 patents in the field of photonics, primarily targeted at increasing capacity, reach and functionality in the optical layer.

Prof. Ellis is a Fellow of Optica. He served for 6 years as an associate editor of the journal Optics Express. He has twice been a Technical Program Committee PC member for OFC. Prof Ellis was also a member of the Technical Program Committee of ECOC from 2004 to 2018 serving as overall TPC chair for ECOC 2019.
\end{IEEEbiography}




\end{document}